\RequirePackage{fix-cm}
\documentclass[smallextended]{svjour3}       
\smartqed  
\usepackage{graphicx}
\usepackage{natbib}
\usepackage{tabularx} 

\usepackage[colorlinks=true, linkcolor=blue, citecolor=blue, urlcolor=blue]{hyperref}
\usepackage{color}

\usepackage{subcaption}
\usepackage{tikz}
\usetikzlibrary{shadows, arrows.meta, positioning}
\usepackage{fontawesome5} 
\usepackage{graphicx,amssymb,amsmath,mathrsfs,bbm}
\usepackage{booktabs}
\usepackage{tcolorbox}
\usepackage{orcidlink}
\DeclareUnicodeCharacter{2264}{\leq}
\DeclareUnicodeCharacter{2265}{\geq}
\begin{document}

\title{When Data Quality Issues Collide: A Large-Scale Empirical Study of Co-Occurring Data Quality Issues in Software Defect Prediction 
}

\titlerunning{When Data Quality Issues Collide}        

\author{Emmanuel Charleson Dapaah\orcidlink{0009-0005-5374-311X} \and Jens Grabowski\orcidlink{0000-0003-2994-3531}}


\institute{Emmanuel C. Dapaah \at
              \email{dapaah@cs.uni-goettingen.de}           
           \and
           Jens Grabowski \at
           \email{grabowski@informatik.uni-goettingen.de}
}

\date{Received: date / Accepted: date}

\maketitle

\begin{abstract}

Software Defect Prediction (SDP) models are central to proactive software quality assurance, yet their effectiveness is often constrained by the quality of available datasets. Prior research has typically examined single issues—such as class imbalance or feature irrelevance—in isolation, overlooking that real-world data problems frequently co-occur and interact. This study presents, to our knowledge, the first large-scale empirical analysis in SDP that simultaneously examines five co-occurring data quality issues—class imbalance, class overlap, irrelevant features, attribute noise, and outliers—across 374 datasets and five classifiers. We employ Explainable Boosting Machines together with stratified interaction analysis to quantify both direct and conditional effects under default hyperparameter settings, reflecting practical baseline usage.

Our results show that co-occurrence is nearly universal: even the least frequent issue (attribute noise) appears alongside others in more than 93\% of datasets. Irrelevant features and imbalance are nearly ubiquitous, while class overlap is the most consistently harmful issue. We identify stable tipping points—around 0.20 for class overlap, 0.65–0.70 for imbalance, and 0.94 for irrelevance—beyond which most models begin to degrade. We also uncover counterintuitive patterns, such as outliers improving performance when irrelevant features are low, underscoring the importance of context-aware evaluation. Finally, we expose a performance–robustness trade-off: no single learner dominates under all conditions.

By jointly analyzing prevalence, co-occurrence, thresholds, and conditional effects, our study directly addresses a persistent gap in SDP research—moving beyond isolated analyses to provide a holistic, data-aware understanding of how quality issues shape model performance in real-world settings.

\end{abstract}

\section{Introduction}
\label{intro}

Software Defect Prediction (SDP) stands as a critical research domain within empirical software engineering and has evolved significantly since its inception in the 1970s with early work by~\cite{halstead_1997} and~\cite{mccabe_1976} on software complexity metrics. Its primary objective is to proactively identify software modules prone to defects before deployment, thereby enabling more targeted testing efforts and optimal resource allocation. The field gained substantial momentum with the pioneering work of~\cite{fenton_quantitative_2000}, who established the theoretical foundations for using statistical models in defect prediction.~\cite{menzies_data_2007} further advanced the field by demonstrating the effectiveness of machine learning (ML) approaches, particularly Naive Bayes classifiers, in analysing historical software metrics to predict the likelihood of defects residing within various software components, establishing a benchmark that influenced subsequent research directions.

Despite significant advancements in developing sophisticated modelling techniques, the practical effectiveness of SDP models remains fundamentally constrained by the inherent quality of the datasets used for their training and validation~\citet{bhandari_data_2023}. Real-world software defect prediction datasets are intrinsically noisy and often plagued by various data quality issues. These issues can profoundly degrade model performance, manifesting in several forms such as severe class imbalance, irrelevant features, and noisy or ambiguous instances. Numerous studies in software engineering and related ML domains have demonstrated that such data quality issues can substantially degrade model performance~\citep{barry_impact_2023,bhandari_data_2023,shivaji_reducing_2013,song_comprehensive_2019}. For instance,~\cite{song_comprehensive_2019} conducted a comprehensive experiment involving 27 datasets, 7 classifiers, and 17 imbalance handling methods, and confirmed that class imbalance markedly degrades default classifier performance. They found that standard learners struggle even with moderate imbalance ratios and recommended using specialized imbalance learning techniques (e.g. resampling or cost-sensitive methods) when the imbalance level is moderate or higher. A study by~\cite{shivaji_reducing_2013} also demonstrated that high-dimensional feature spaces, common in software metrics datasets, can lead to overfitting and reduced generalization capability.

While the crucial importance of data quality in SDP is widely acknowledged, a notable limitation persists: existing research predominantly investigates data quality issues in isolation. Most studies typically focus on the impact of a single quality issue, such as class imbalance or noise, while either controlling for or ignoring other concurrent issues.  This reductionist approach, although methodologically sound for understanding individual effects, falls short of capturing the intricate reality of software defect datasets. In practice, multiple quality issues frequently coexist and can interact in complex ways. For example, a dataset exhibiting severe class imbalance might simultaneously suffer from high irrelevant features and class overlap. Such concurrent conditions could collectively amplify, dampen, or even reverse the effect of any single issue, highlighting the necessity of adopting a more holistic perspective on data quality in SDP research.

A further significant limitation is the scarcity of empirical studies that thoroughly examine the impact of less-explored issues such as class overlap, attribute noise, and outliers. Although these factors are frequently presumed to be influential, they largely lack comprehensive empirical validation within the context of SDP. Additionally, there is a significant gap in systematically exploring how various machine learning paradigms respond to multifaceted data quality conditions. This knowledge gap impedes practitioners’ ability to confidently select models that are robust and effective given specific dataset characteristics, particularly when confronted with the complex interplay of multiple data quality issues.

To address these limitations, this study presents a large-scale systematic empirical investigation of the impact of data quality issues on SDP model performance, explicitly accounting for the concurrent presence of multiple quality issues. Our approach moves beyond the traditional single-issue analysis paradigm to examine how individual data quality issues influence model performance relative to the presence and severity of other quality issues. Specifically, we investigate five critical data quality issues—\textit{Class Imbalance}, \textit{Class Overlap}, \textit{Irrelevant features}, \textit{Attribute Noise}, and \textit{Outliers}—across five widely adopted machine learning algorithms in software defect prediction: \textit{Decision Trees}, \textit{Random Forest}, \textit{Naive Bayes}, \textit{Multilayer Perceptron}, and \textit{Support Vector Machines}. Our analysis encompasses 374 diverse software defect prediction datasets, providing unprecedented scope for understanding the influence of these data quality issues.

This study is guided by the following research questions (RQs):
\begin{itemize}
    \item \textbf{RQ1}: How prevalent are various data quality issues across a large, diverse collection of SDP datasets?
    \item \textbf{RQ2}: What is the baseline performance of common machine learning classifiers on SDP datasets with co-occurring quality issues?
    \item \textbf{RQ3}: What is the relative influence and monotonic relationship of each data quality issue on SDP model performance, and how do these patterns change in the presence of other co-occurring issues?
\end{itemize} 

This study offers several contributions that advance the state of empirical research in SDP:

\begin{enumerate}
    \item We conduct, to our knowledge, the first large-scale empirical study in SDP that jointly evaluates five key data quality issues—including three underexplored ones (class overlap, attribute noise, and outliers)—across 374 datasets and five widely used classifiers, using Explainable Boosting Machines (EBMs) for interpretable analysis.

    \item We provide systematic, large-scale evidence that data quality issues co-occur almost universally in SDP datasets, showing that multi-problem conditions are the norm rather than isolated cases.

    \item We employ Explainable Boosting Machines (EBMs) and stratified interaction analysis to quantify both direct and conditional effects, enabling interpretable modeling of complex dependencies.

    \item We uncover stable thresholds—around 0.20 (overlap), 0.65–0.70 (imbalance), and 0.94 (irrelevance)—beyond which model performance consistently deteriorates, offering actionable guidance.

    \item We identify counterintuitive effects (e.g., outliers improving performance under certain conditions), challenging prevailing preprocessing practices such as automatic outlier removal.

    \item We expose a performance–robustness trade-off across learners, showing that no single model is universally resilient across data quality profiles.

    \item We release all datasets and analysis scripts to support open, reproducible research \protect\footnotemark.
\end{enumerate}

\footnotetext{https://github.com/ecdapaah-dev/When-Data-Issues-Collide.git}

The remainder of this paper is organized as follows. Section \ref{related_work} reviews related work on individual and co-occurring data quality issues in SDP. Section~\ref{methodology} details our methodology, including dataset selection, quality quantification, and modeling approach. Section~\ref{results} presents the empirical results in response to our research questions. Section~\ref{implications} discusses implications for researchers and practitioners. Section \ref{threats_to_validity} outlines threats to validity and Section~\ref{conclusion} concludes the paper with directions for future research.

\section{Related Work}
\label{related_work}
This section reviews prior research on data quality issues in SDP, organized into two key strands: (i) studies that examine individual data quality issues in isolation and (ii) efforts that attempt to address or model multiple co-occurring issues. While the former dominates the literature, the latter—though more realistic—remains underdeveloped.

\subsection{Studies on Individual Data Quality Issues}

Irrelevant features—software metrics that contribute little or no predictive value to the target variable—increase computational overhead, degrade predictive performance, and heighten the risk of overfitting. Research addressing this issue commonly employs feature selection techniques to identify optimal subsets of informative features. For example,~\cite{wang_software_2012} explored ensemble techniques for data reduction, finding that small ensembles of feature rankers are highly effective.~\cite{chen_empirical_2013} proposed a two-stage framework for feature selection addressing relevance and redundancy.~\cite{kumar_effective_2018} used an Extreme Learning Machine with feature selection to address the challenges of increased model complexity and longer execution times often associated with high dimensionality.~\cite{jian_hybrid_2019} introduced a hybrid feature selection method combining clustering with wrapper-based selection to filter out irrelevant and redundant features.~\cite{borandag_majority_2018} developed a Majority Vote based Feature Selection algorithm to identify the most valuable software metrics for fault prediction. These studies affirm a direct link between irrelevant features and performance degradation, coupled with increased computational burden. However, a critical observation is the implicit assumption of orthogonality, where studies address this data quality issue in isolation, potentially overlooking synergistic or antagonistic effects with other data quality issues.

Class imbalance, where one class (e.g., defective modules) is underrepresented, biases machine learning models towards the majority class, leading to poor prediction for the minority.~\cite{tan_online_2015} focused on online defect prediction for imbalanced data, using resampling and updatable classification. They identified that data imbalance, characterized by significantly fewer buggy changes than clean changes, contributes to low prediction performance.~\cite{li_oversampling_2016} proposed an oversampling-boosting technique that generates synthetic samples to correct the bias of SDP models toward the majority class. Similarly,~\cite{bejjanki_class_2020} introduced the Class Imbalance Reduction method, which generates synthetic instances around the centroids and nearest neighbors of the minority class to address prediction bias in imbalanced datasets.~\cite{liu_comparative_2022} conducted a comparative study on data imbalance, finding that sampling techniques significantly impact defect prediction outcomes. These studies demonstrate that imbalanced data directly compromises how machine learning models learn, resulting in suboptimal performance, particularly for the critical minority class. A critical consideration is that solutions for class imbalance, such as oversampling, can inadvertently introduce or exacerbate other issues like class overlap and noise. This underscores the intricate interdependencies among data quality issues and reveals the limitations of examining them in isolation.

Research on noise in SDP has been relatively limited, with most studies concentrating on label noise.~\cite{khoshgoftaar_generating_2004} investigated the use of ensemble-partitioning filters to generate multiple noise elimination models aimed at removing noisy instances and improving SDP model performance.~\cite{catal_class_2011} proposed a class noise detection approach based on threshold values derived from software metrics and ROC curve analysis, aiming to identify and eliminate mislabeled instances that degrade predictive accuracy.~\cite{khan_impact_2021} examined the impact of label noise and evaluated the effectiveness of various noise filters, highlighting the significant influence of label noise on model performance. Despite these valuable contributions, the literature on noise remains fragmented and rarely addresses its interdependency with other data quality issues.

Outliers are data points that deviate significantly from the norm and can distort statistical analyses and adversely affect model training. In the context of SDP, outlier detection has received moderate attention, with most approaches relying on threshold-based techniques.~\cite{alan_outlier_2009} developed early outlier detection algorithms using thresholds derived from object-oriented software metrics, establishing baseline methods for identifying anomalous software modules. In a subsequent study,~\cite{alan_thresholds_2011} refined this approach by focusing on class outliers, which are instances whose labels are inconsistent with their feature values. However, a notable theme in outlier research is its conceptual and practical overlap with noise. As noted by~\cite{samami_mixed_2020}, "potential noisy samples may include outliers," suggesting an interdependency between these two data quality issues.

Class overlap occurs when instances from different classes share similar feature values, hindering classifier decision boundaries and potentially degrading model performance more than class imbalance.~\cite{feng_roct_2021} proposed ROCT (Radius-based Class Overlap Cleaning Technique) to alleviate class overlap in software defect prediction. The core contribution of ROCT is its ability to identify and remove overlapping instances by treating each instance as the center of a hypersphere and optimizing its radius, crucially operating without requiring hyperparameter tuning.~\cite{gong_comprehensive_2023} conducted a comprehensive investigation of the impact of class overlap on software defect prediction. They observed that 70\% of SDP datasets contain overlapping instances and that different levels of class overlap have varying impacts on model performance. 
Although these studies provide foundational evidence of the problem, they do not investigate how class overlap may be influenced by co-occurring data quality issues—a gap that this study directly addresses.

\subsection{Attempts at Handling Multiple Issues}
While the majority of SDP data quality research has focused on individual issues, a smaller but growing body of work has begun to explore integrated or hybrid approaches aimed at addressing multiple co-occurring data quality problems. These studies acknowledge that real-world datasets rarely exhibit a single isolated issue.

\cite{cao_software_2025} proposed FC-BLS, a Fuzzy Cost Broad Learning System designed to enhance defect prediction by jointly addressing class imbalance, outliers, and noise.~\cite{gong_tackling_2019} addressed class imbalance through a cluster-based oversampling technique combined with noise filtering, thereby integrating rebalancing and data cleaning strategies.~\cite{joon_noise_2020} implemented a comprehensive pipeline that incorporates noise filtering, class imbalance correction, and irrelevant feature removal.

While these approaches represent a critical shift toward more realistic modeling, they share an important limitation: they do not disentangle the individual contributions of each data quality issue to model performance, nor do they examine how such issues condition each other’s effects.

In sum, the literature confirms that individual data quality issues—such as class imbalance, irrelevant features, and noise—negatively affect SDP models. However, most studies assume independence between these issues and do not empirically examine their joint or conditional effects. Moreover, underexplored factors like class overlap and outliers remain inadequately validated in empirical contexts. To address these limitations, we move beyond single-issue analyses by systematically examining five co-occurring data quality issues in SDP across 374 datasets and five classifiers. Building on EBMs and stratified interaction analysis, our study provides a holistic view of how data quality profiles shape model performance

\section{Methodology}
\label{methodology}
Having established the need for a holistic investigation of co-occurring data quality issues, we now present our methodological framework. This covers dataset selection and preprocessing, formal quantification of the five quality issues, model configuration, evaluation metrics, and the analytical techniques used to disentangle direct and conditional effects.

\subsection{Datasets and Preprocessing}

We curated a large and diverse benchmark of SDP datasets drawn from three established public repositories: the GitHub Bug Dataset~\citep{toth_public_2016}, the Large Defect Prediction benchmark~\citep{tantithamthavorn_large_2022}, and the Unified Bug Dataset~\citep{ferenc_public_2018}. These repositories span open-source projects, multiple programming languages, and varied development practices, providing rich heterogeneity in data quality characteristics.

To ensure consistency across heterogeneous SDP datasets and maintain the validity of our analysis, we implemented a standardized preprocessing pipeline comprising the following steps:

\begin{enumerate}
    \item \textbf{Format Standardization:} All datasets were converted to a common CSV format for uniform processing.
    
    \item \textbf{Release Filtering:} Multiple releases of the same dataset were retained only when they exhibited substantial changes in the predictive feature set (e.g., introduction of new software metrics or significant modifications to existing ones). This criterion ensured that successive releases represented meaningful variations in data characteristics rather than minor updates, thereby preventing artificial inflation of the dataset pool.
    
    \item \textbf{Binary Label Harmonization:} Many datasets used different labeling schemes, such as numeric defect counts, severity levels, or Boolean flags. To ensure consistency, we reformulated all tasks as binary classification problems by mapping class labels into two categories: defective (modules with one or more reported defects) and non-defective (modules with zero reported defects). For datasets with defect severity or categorical labels, all non-zero or “buggy” indicators were aggregated under the defective class.
    
     \item \textbf{Identifier and Duplicate Column Removal:} Non-predictive identifier columns (e.g., file names, version IDs) were removed, and duplicate columns arising from dataset merging or formatting inconsistencies were eliminated.

    \item \textbf{Missing Data Handling:}
    \begin{itemize}
        \item Instances with missing target labels were removed to preserve validity in supervised learning.
        \item Feature columns with more than 40\% missing values were discarded. Retaining such features would require extensive imputation, which can distort the intrinsic structure and complexity of the dataset---an essential factor for this study, which aims to assess the natural influence of data quality issues on model performance. Removing these features ensures that the observed effects reflect genuine data quality characteristics rather than artifacts of synthetic imputation \citep{garcia_2010}.
        \item For remaining numeric features, missing values were imputed using median imputation, which is robust to outliers and skewed distributions while maintaining central tendency.
        \item Categorical features were label-encoded after imputation to ensure compatibility with machine learning models.
    \end{itemize}

    \item \textbf{Minimum Data Sufficiency:} To ensure valid stratified train-test splits and maintain reliable performance estimates, we enforced two conditions: (i) datasets with fewer than 100 instances were excluded, and (ii) datasets where either class contained fewer than two instances were discarded. The 100-instance threshold was chosen to balance statistical stability with dataset diversity. Raising this threshold beyond 100 would eliminate a significant number of real-world datasets, thereby reducing coverage and compromising the external validity of our study. These criteria guarantee both overall size adequacy and minimal class representation, reducing variance and ensuring stability in model evaluation.

\end{enumerate}

These processes allowed us to retain datasets with naturally occurring quality issues, while filtering out those that were structurally unusable for our analysis. In total, 374 datasets were included in the study, offering a broad and realistic spectrum of data quality profiles. Table~\ref{dataset} summarizes the number of datasets available and selected within each benchmark group.

\begin{table}[ht]
\centering
\caption{Experimental Datasets}
\label{dataset}
\begin{tabular}{llll}
\toprule
\textbf{Benchmark} & \textbf{Dataset Group} & \textbf{Total} & \textbf{Selected} \\
\midrule
GitHub Bug Dataset\citep{toth_public_2016} & antlr4 & 10 & 10 \\
 & BroadleafCommerce & 22 & 10 \\
 & ceylon-ide-eclipse & 10 & 4 \\
 & elasticsearch & 24 & 23 \\
 & hazelcast & 18 & 16 \\
 & junit & 16 & 10 \\
 & MapDB & 12 & 10 \\
 & mcMMO & 12 & 8 \\
 & mct & 6 & 6 \\
 & neo4j & 18 & 12 \\
 & netty & 18 & 16 \\
 & orientdb & 12 & 10 \\
 & oryx & 8 & 6 \\
 & titan & 12 & 10 \\
\midrule
Large Defect Prediction\citep{tantithamthavorn_large_2022} & AEEEM & 5 & 5 \\
 & JIRA & 32 & 32 \\
 & JIT & 2 & 2 \\
 & NASA & 39 & 37 \\
 & Relink & 5 & 4 \\
 & TeraPromise & 88 & 65 \\
\midrule
Unified Bug Dataset\citep{ferenc_public_2018} & Bugcatchers & 3 & 3 \\
 & BugPrediction & 5 & 5 \\
 & GitHub & 30 & 28 \\
 & PROMISE & 45 & 39 \\
 & Zimmerman & 3 & 3 \\
\bottomrule
\end{tabular}
\end{table}

\subsection{Measuring Data Quality Issues}

We operationalized five distinct data quality issues using established information-theoretic and statistical formulations: \textit{irrelevant features}, \textit{class imbalance}, \textit{class overlap}, \textit{attribute noise}, and \textit{outliers}. Each issue was quantified for every dataset in our benchmark and subsequently used as an independent variable in the modeling process. The formal definitions and computation methods for each quality issue are detailed below.

\paragraph{\textbf{Irrelevant Features}} pertains to datasets in which a significant number of features contribute little to no predictive information about the target variable, introducing noise and increasing the risk of overfitting. To quantify this, we compute the mutual information (MI) between each feature $x$ and the target variable $y$, using the standard formulation~\citep{michie_1995}:

\begin{equation}
\text{MI}(x, y) = H(x) + H(y) - H(x, y)
\end{equation}

where $H(\cdot)$ denotes the entropy function. \(\text{MI}(x, y)\) reflects the amount of information shared between a feature and the target—higher values indicate stronger predictive relevance. To align interpretation with other data quality metrics (where higher values signify poorer quality), we invert this measure by computing $1 - \text{MI}(x, y)$. This transformation ensures that higher scores consistently indicate greater irrelevance or redundancy, thereby promoting interpretive uniformity across all quality issues.

\paragraph{\textbf{Class Imbalance}} captures the disproportional distribution of instances across classes, which often leads to biased learning and degraded predictive performance. In the context of SDP, it typically reflects the underrepresentation of defect-prone modules. We adopt the imbalance measure from~\citep{lorena2020}, a normalized imbalance ratio suitable for both binary and multiclass classification. It is defined as:

\begin{equation}
Class\ Imbalance = 1 - \frac{1}{\text{IR}}
\end{equation}

where the imbalance ratio IR is computed as:

\begin{equation}
\text{IR} = \frac{n_c - 1}{n_c} \sum_{i=1}^{n_c} \frac{n_i}{n - n_i}
\end{equation}

Here, \textit{n\textsubscript{c}} is the number of classes, \textit{n\textsubscript{i}} is the number of instances in class \textit{i}, and \textit{n} is the total number of instances in the dataset. Values of the class imbalance measure range from 0 (perfectly balanced) to 1 (highly imbalanced), making it a standardized and interpretable indicator of class imbalance severity.

\paragraph{\textbf{Class Overlap}} represents the extent to which instances from different classes are interspersed in the feature space, which increases classification complexity. To capture this phenomenon, we compute overlap based on a Minimum Spanning Tree (MST) constructed from all instances, where nodes correspond to examples and edge weights are given by their pairwise distances. The overlap score is defined as the fraction of edges in the MST that connect instances from different classes~\citep{lorena2020}:

\begin{equation}
\text{Overlap} = \frac{1}{n} \sum_{i=1}^{n} \mathbb{I}\big( (x_i, x_j) \in \text{MST} \land y_i \neq y_j \big)
\end{equation}

Here, \(n\) is the number of instances, \((x_i, x_j)\) represents an edge in the MST, \(y_i\) and \(y_j\) are the corresponding class labels, and \(\mathbb{I}(\cdot)\) is an indicator function. The values of the class overlap measure lie in the range [0, 1], where higher values indicate greater class overlap.

\paragraph{\textbf{Outliers}} refer to values that significantly deviate from the majority of the data and may arise due to anomalies, rare conditions, or measurement noise. We estimate outliers using the Interquartile Range (IQR) method~\citep{tukey_1977}. For each numeric feature, the first (\(Q_1\)) and third (\(Q_3\)) quartiles are computed, and the IQR is defined as \(\text{IQR} = Q_3 - Q_1\). A value is flagged as an outlier if it falls outside the interval:

\begin{equation}
[Q_1 - w \cdot \text{IQR}, \; Q_3 + w \cdot \text{IQR}]
\end{equation}

where \(w\) is typically set to 1.5. To enable comparability across datasets of different sizes and dimensionalities, we normalize by the total number of values, defining:

\begin{equation}
\text{Outlier Ratio} = \frac{\text{\# Outlier Values}}{n \times m}
\end{equation}

where \(n\) is the number of instances and \(m\) the number of features. Higher values indicate a greater proportion of outlier values, reflecting increased irregularity in the dataset.

\paragraph{\textbf{Attribute Noise}} captures the degree to which feature values exhibit non-informative or random variation with respect to the target variable. We compute this using a ratio-based measure that compares the total entropy of all predictive attributes to the amount of information they provide about the class label~\citep{michie_1995}. Specifically, the noise score N is defined as:

\begin{equation}
N = \frac{\sum_{x} H(x) - \sum_{x} \text{MI}(x, y)}{\sum_{x} \text{MI}(x, y)}
\end{equation}

where H(x) is the entropy of attribute \textit{x}, and MI(x,y) is the mutual information between attribute \textit{x} and the target \textit{y}. Both summations are computed across all features in the dataset. A higher value of N indicates that the attributes collectively carry a large amount of entropy (uncertainty) that is not informative for predicting the target, suggesting a greater presence of noise. For comparability across datasets, \(N\) is min–max normalized to the range \([0, 1]\), ensuring that higher normalized values consistently reflect higher relative noise levels.

\subsection{Models and Configurations}
We selected five widely adopted classifiers in SDP literature~\citep{bhandari_data_2023}, each representing a distinct learning paradigm: tree-based, ensemble, kernel-based, probabilistic, and neural architectures.

\begin{itemize}
    \item \textbf{Decision Tree (DT)}:  
    a non-parametric, tree-based model that recursively partitions the data using axis-aligned splits on feature values, yielding an interpretable hierarchy of decision rules~\citep{quinlan_induction_1986}.

    \item \textbf{Random Forest (RF)}:  
    an ensemble method that constructs a collection of decision trees on bootstrapped subsets of the training data and aggregates their predictions. RFs improve robustness and generalization compared to single trees by mitigating variance through random feature selection~\citep{breiman_random_2001}.

    \item \textbf{Support Vector Machine (SVM)}:  
    a margin-based classifier that identifies the optimal separating hyperplane between classes in a transformed feature space. SVMs use support vectors (the closest data points to the decision boundary) to define this hyperplane and can handle non-linear classification through kernel functions that implicitly map data to higher-dimensional feature spaces~\citep{cortes_1995}.

    \item \textbf{Gaussian Naive Bayes (NB)}:  
    a probabilistic model grounded in Bayes’ Theorem that assumes conditional independence among features~\citep{david_2001}. NB models the distribution of continuous features using Gaussian likelihoods and is valued for its simplicity and efficiency.

    \item \textbf{Multi-Layer Perceptron (MLP)}:  
    a feedforward neural network composed of fully connected layers, trained via backpropagation~\citep{rumelhart_learning_1986}. MLPs are capable of learning complex, non-linear relationships but require sufficient data and computational resources to avoid overfitting.
\end{itemize}

\paragraph{\textbf{Rationale for Default Hyperparameters:}} 
All models were evaluated using default hyperparameter configurations from \texttt{scikit-learn}~\citep{scikit-learn}. 
This design choice reflects two motivations: (i) defaults provide a practical and widely used 
baseline in SDP research and practice, and (ii) they ensure fair comparison by avoiding selective 
tuning that could bias results toward certain models. However, this also means that some models 
known to benefit strongly from tuning---in particular, SVM and MLP---may appear comparatively 
weaker, while tree-based models (DT, RF) may seem stronger under default settings. 
As a result, the performance--robustness trade-off observed in our results should be interpreted 
as representative of baseline usage rather than the absolute performance ceiling of each learner.

\subsection{Evaluation Metric}

To evaluate the predictive performance of each model across diverse SDP datasets with varying class distributions, we employed \textit{Balanced Accuracy} as the primary performance metric. Balanced Accuracy is particularly well-suited for imbalanced classification tasks, as it equally weights performance on both the majority and minority classes—an important consideration in software defect prediction, where non-defective modules typically outnumber defective ones.

For binary classification, Balanced Accuracy~\citep{brodersen_2010} is computed as the average of sensitivity (true positive rate) and specificity (true negative rate):

\begin{equation}
\text{Balanced Accuracy} = \frac{1}{2} \left( \frac{TP}{TP + FN} + \frac{TN}{TN + FP} \right)
\end{equation}

Where:
\begin{itemize}
    \item \textit{TP} denotes True Positives (correctly predicted defective modules),
    \item \textit{TN} denotes True Negatives (correctly predicted non-defective modules),
    \item \textit{FP} denotes False Positives (non-defective modules incorrectly classified as defective), and
    \item \textit{FN} denotes False Negatives (defective modules incorrectly classified as non-defective).
\end{itemize}

This metric provides an unbiased evaluation by mitigating the inflated performance scores often produced by accuracy in imbalanced datasets. 

\subsection{Analytical Approach}

\paragraph{\textbf{Explainable Modeling:}} To quantify the effect of each data quality issue on model performance, we employed Explainable Boosting Machines~\citep{nori_2019}. EBMs are a powerful and interpretable machine learning model that provides global and local explanations. Unlike black-box models, EBMs explicitly learn Generalized Additive Models (GAMs) by combining small decision trees. This allows us to observe how each feature independently affects the prediction outcome while accounting for other features. For each classifier, an EBM was trained using the five data quality metrics as input features and the corresponding balanced accuracy as the target. EBMs provide two critical forms of interpretability:

\begin{itemize}
\item \textbf{Feature Influence Scores:} these quantify the relative influence of each data quality issue on model performance.
\item \textbf{Shape functions:} these describe the form and direction (e.g., monotonicity) of the relationship between each quality issue and balanced accuracy. It further enables the identification of non-linear or threshold effects.
\end{itemize}

\paragraph{\textbf{Interaction Analysis:}} To investigate potential conditional dependencies between quality issues, we conduct a series of stratified interaction analyses. For each analysis, we selected a primary data quality issue and stratified the dataset collection into tertiles (low, medium, high) based on the severity of a second, co-occurring issue. Within each stratum, we examined the relationship between the primary issue and balanced accuracy across models. This allowed us to assess whether the direction or strength of influence varied across different conditions, thereby providing insights into possible interaction effects. 

This methodology was particularly useful in interpreting counterintuitive findings from the EBM shape functions and served as an additional layer of diagnostic analysis in our evaluation pipeline.

\section{Results and Discussion}
\label{results}
Building on this methodological foundation, we now present our empirical findings. The results are organized around the three research questions, moving from the prevalence of data quality issues, through baseline model performance under real-world conditions, to the direct and conditional effects of each issue on classifier robustness.

\subsection{Prevalence of Data Quality Issues}
\label{sub1}
\begin{figure}[htbp]
    \centering
    \begin{subfigure}{0.49\textwidth}
        \centering
        \includegraphics[width=\linewidth]{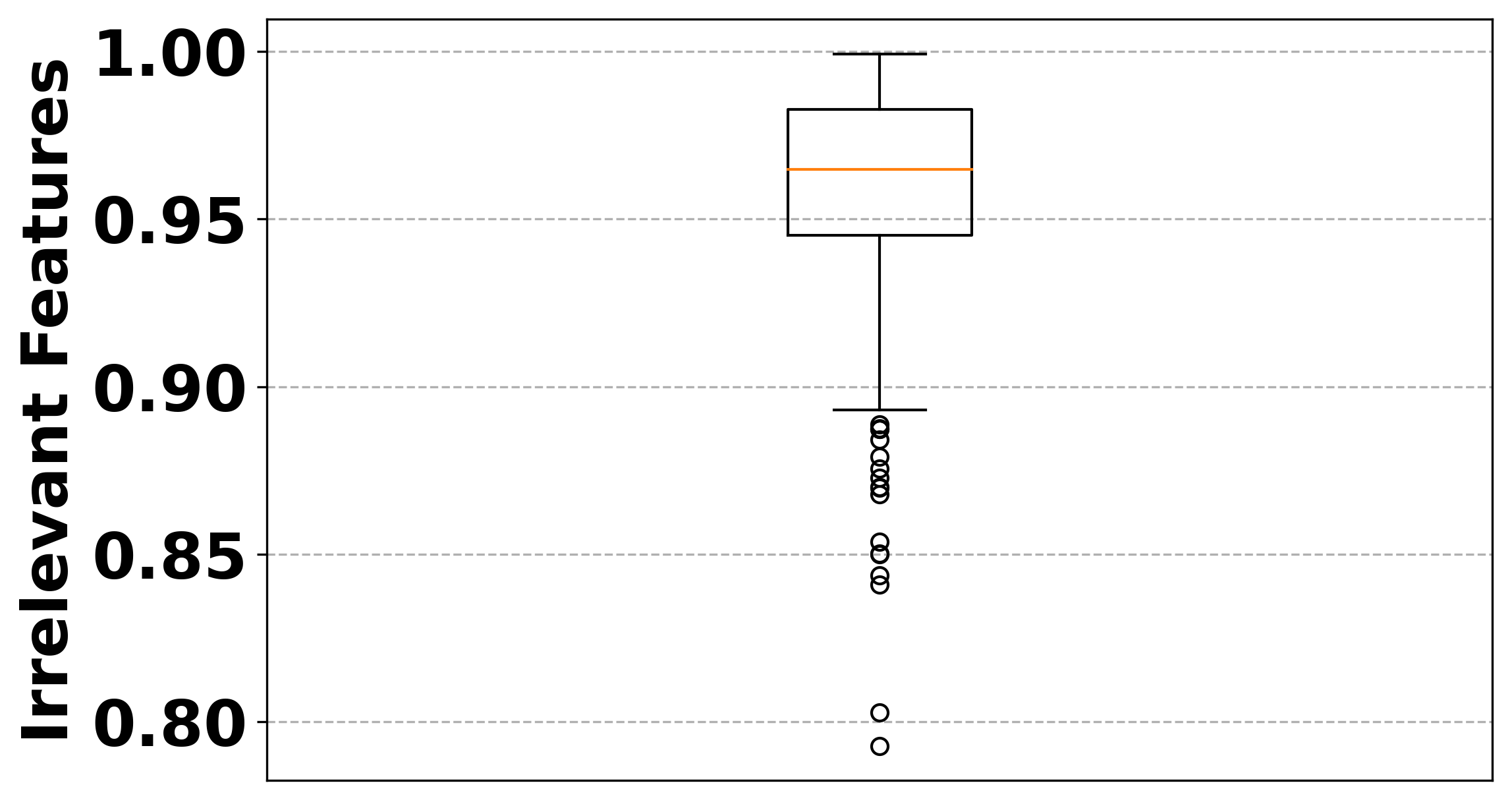}
        \caption{Irrelevant Features}
        \label{boxplot_dimensionality}
    \end{subfigure}
    \begin{subfigure}{0.49\textwidth}
        \centering
        \includegraphics[width=\linewidth]{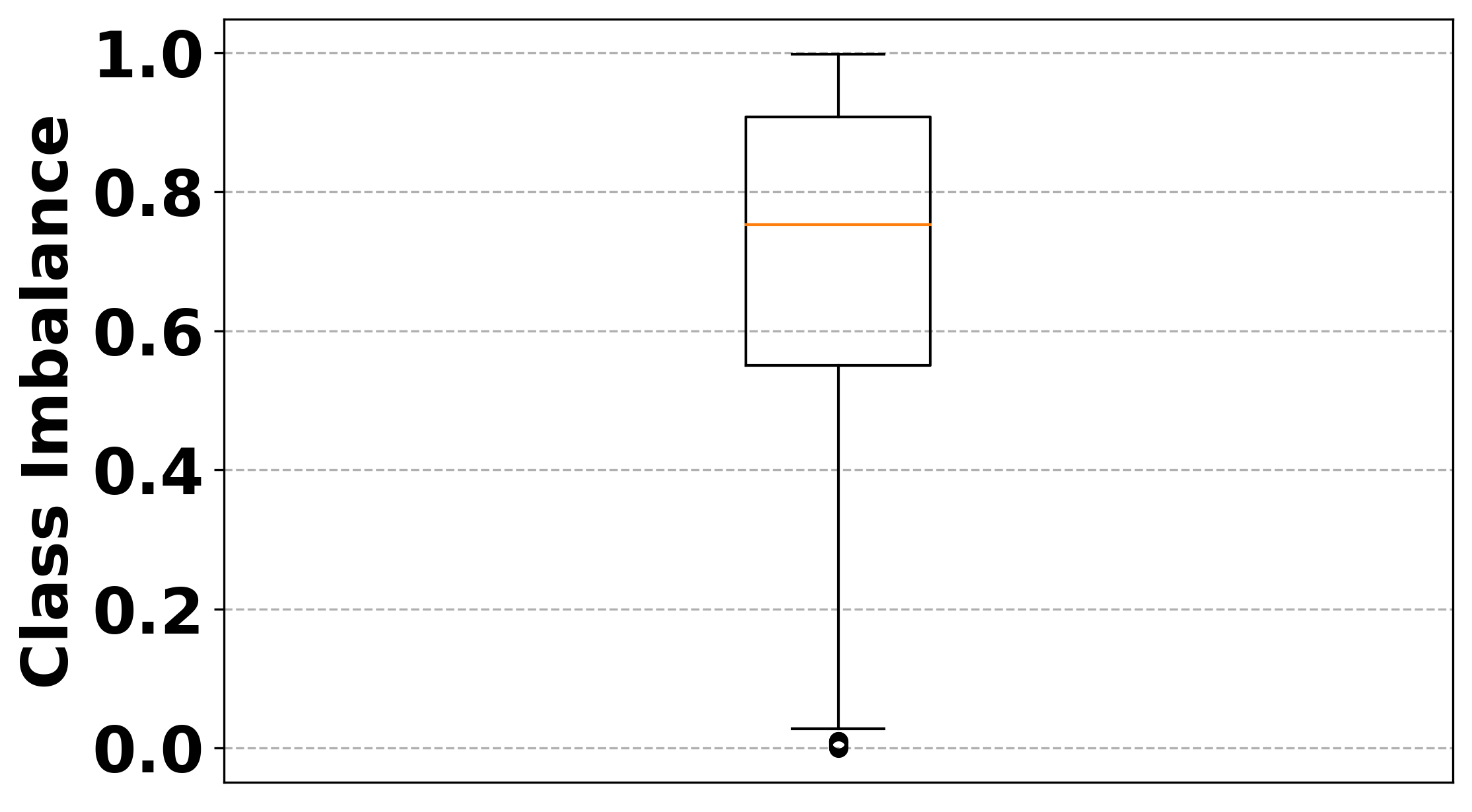}
        \caption{Class Imbalance}
        \label{boxplot_imbalance}
    \end{subfigure}
     \vspace{0.2cm} 
    \begin{subfigure}{0.49\textwidth}
        \centering
        \includegraphics[width=\linewidth]{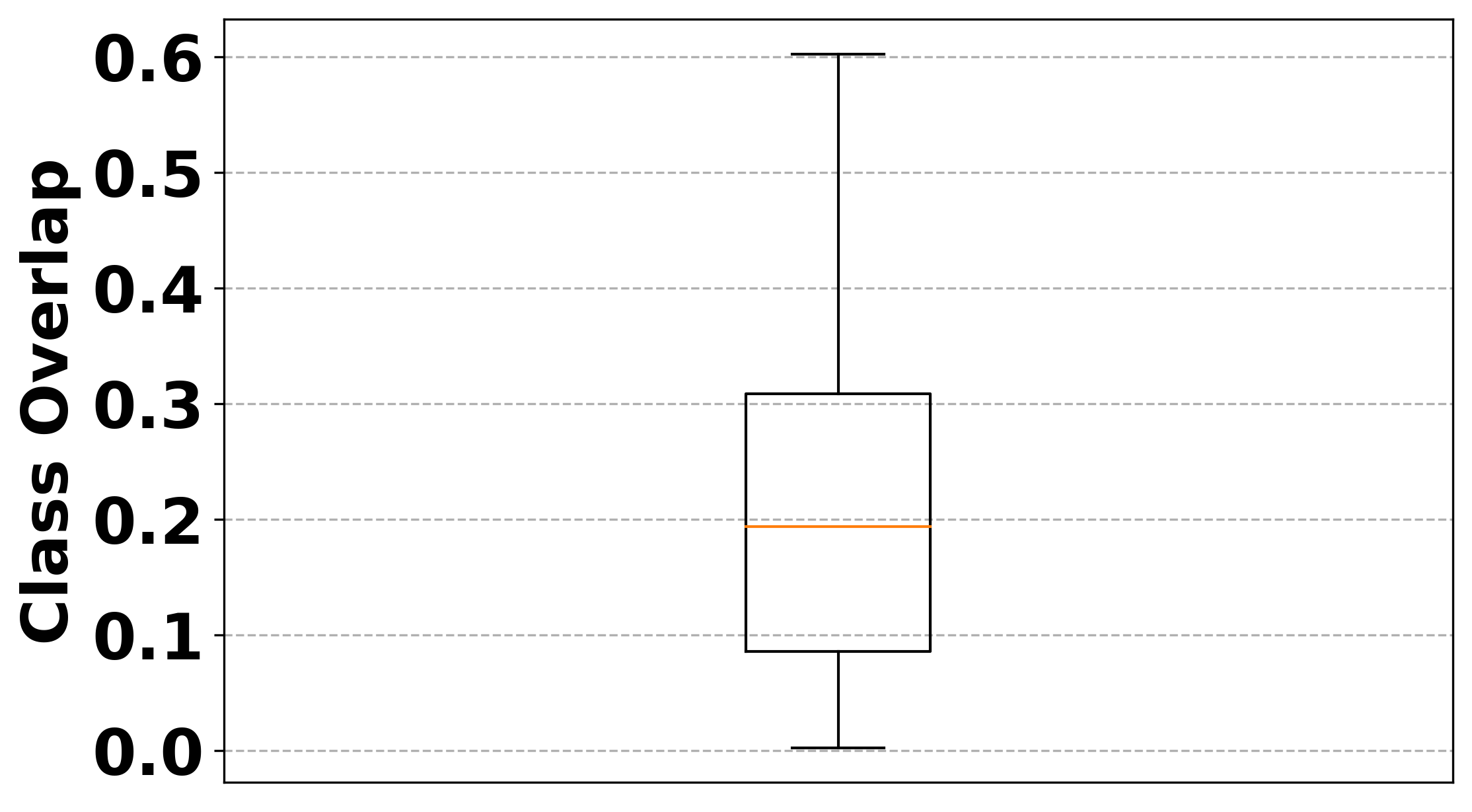}
        \caption{Class Overlap}
        \label{boxplot_overlap}
    \end{subfigure}
    \begin{subfigure}{0.49\textwidth}
        \centering
        \includegraphics[width=\linewidth]{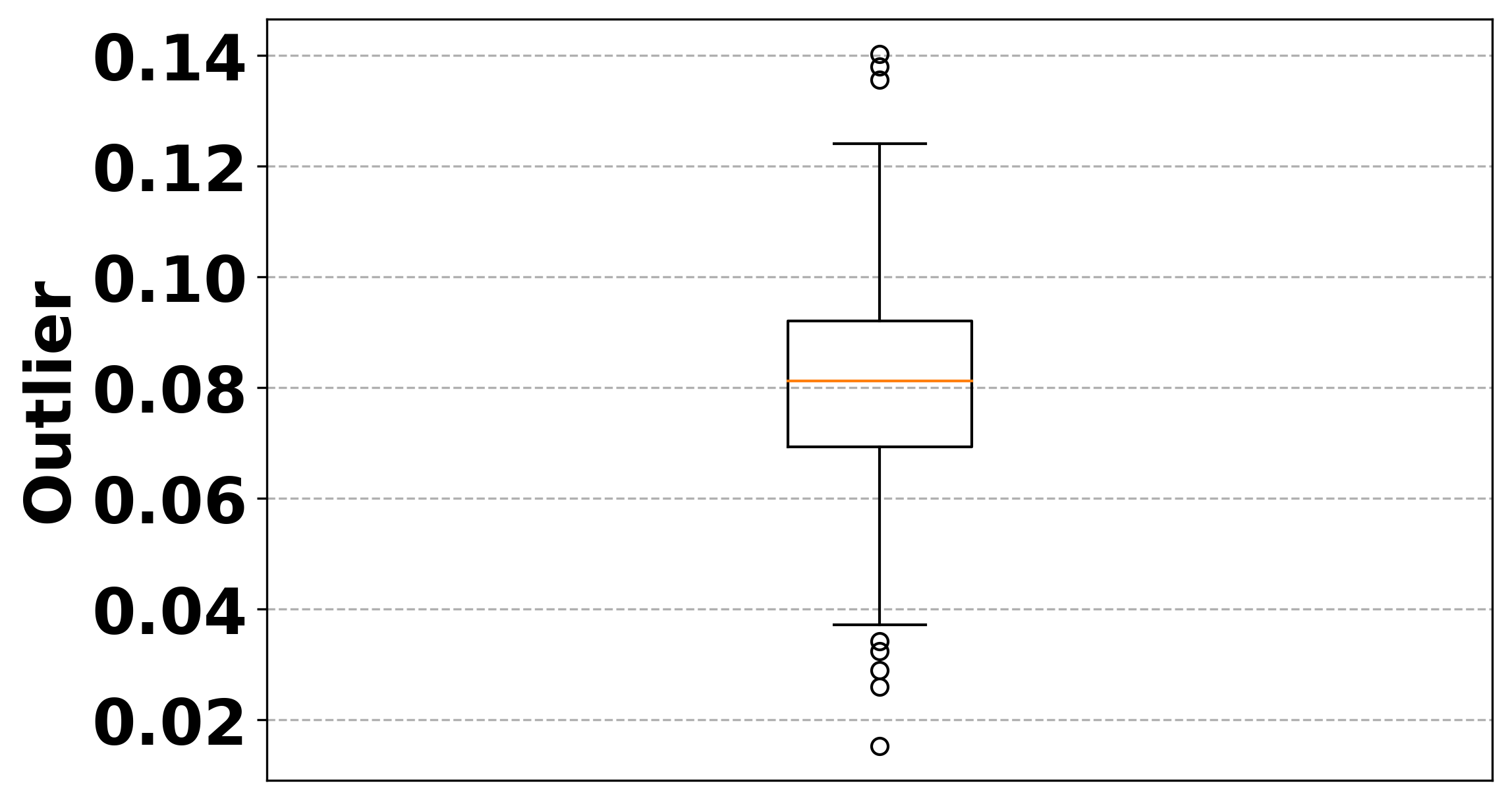}
        \caption{Outlier}
        \label{boxplot_outlier}
    \end{subfigure}
    \vspace{0.2cm} 
    \begin{subfigure}{0.49\textwidth}
        \centering
        \includegraphics[width=\linewidth]{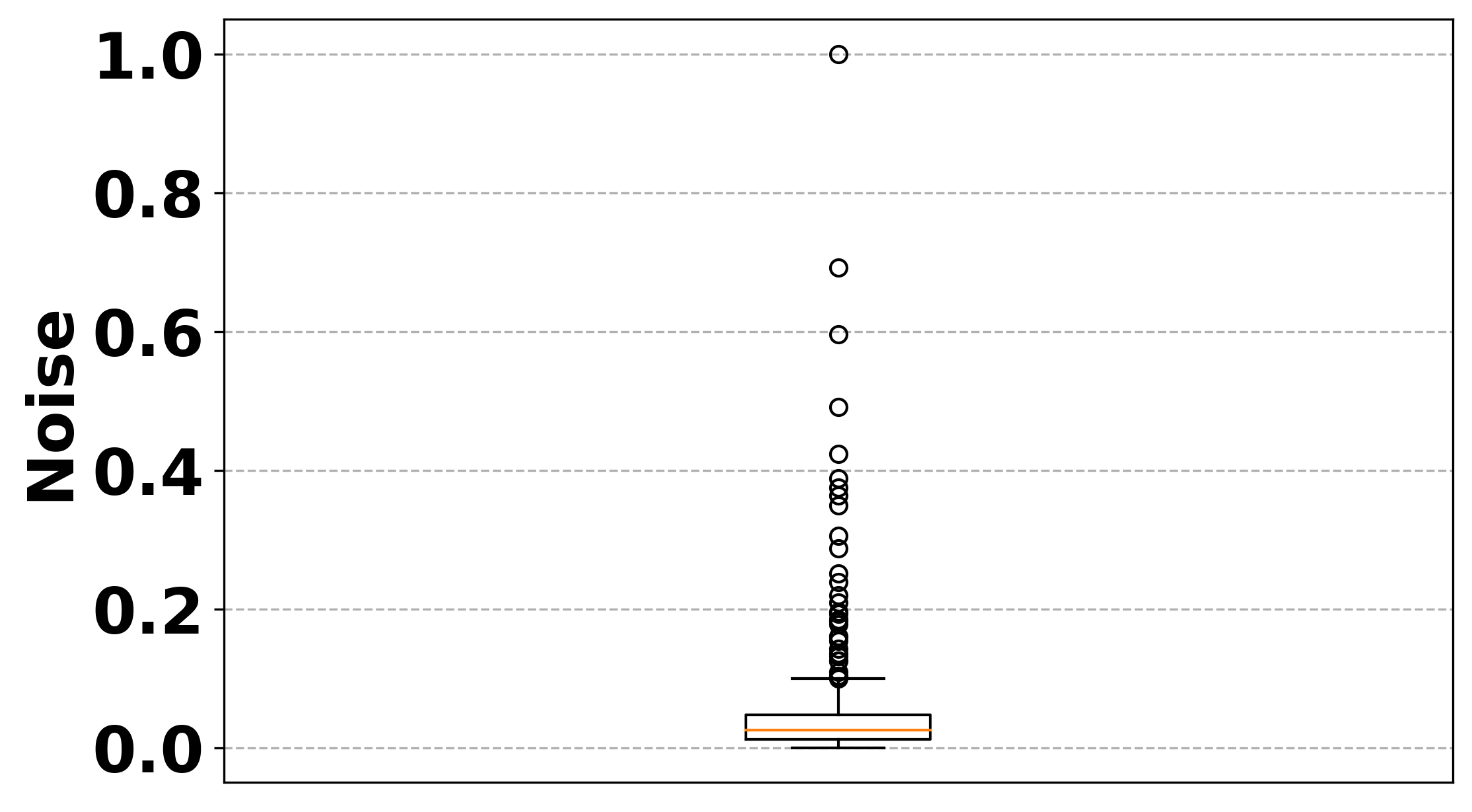}
        \caption{Attribute Noise}
        \label{boxplot_noise}
    \end{subfigure}  

    \caption{Box plots illustrating the distribution of data quality issues}
    \label{data_quality_boxplots}
\end{figure}

\begin{figure}[htbp]
    \centering
    \includegraphics[width=0.9\textwidth]{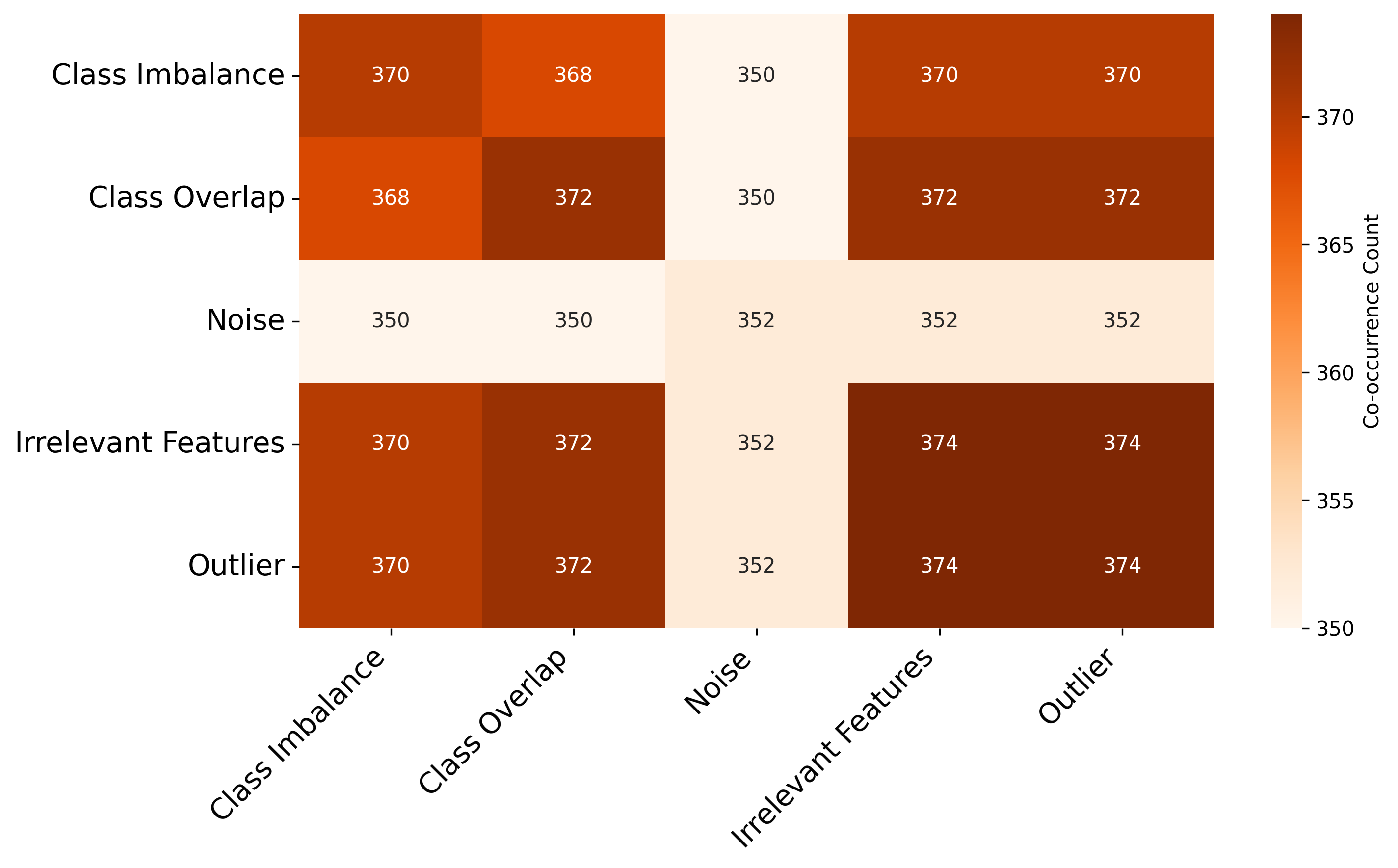}
    \caption{Heatmap illustrating the pairwise co-occurrence of data quality issues}
    \label{co_occurrence_heatmap}
\end{figure}

We begin by addressing RQ1, which asks how prevalent different data quality issues are across SDP datasets. Descriptive statistics (Table~\ref{quality_stats} \& Figure~\ref{data_quality_boxplots}) and co-occurrence patterns (Figure~\ref{co_occurrence_heatmap}) provide the foundation for understanding the conditions under which models are later evaluated.

\begin{table}[ht]
\centering
\caption{Descriptive Statistics of Data Quality Issues}
\begin{tabular}{lcccccc}
\toprule
\textbf{Data Quality Issue} & \textbf{Min} & \textbf{Max} & \textbf{Mean} & \textbf{S.D.} & \textbf{25\%} & \textbf{75\%} \\
\toprule
Irrelevant Features & 0.79 & 0.99 & 0.96 & 0.03 & 0.95 & 0.98 \\
\midrule
Class Imbalance & 0.00 & 0.99 & 0.68 & 0.28 & 0.55 & 0.91 \\
\midrule
Class Overlap & 0.00 & 0.60 & 0.21 & 0.14 & 0.09 & 0.31 \\
\midrule
Outlier & 0.02 & 0.14 & 0.08 & 0.02 & 0.07 & 0.09 \\
\midrule
Attribute Noise & 0.00 & 1.00 & 0.05 & 0.09 & 0.01 & 0.05 \\
\bottomrule
\end{tabular}
\label{quality_stats}
\end{table}

\paragraph{\textit{Irrelevant Features}} emerges as the most pervasive and consistent quality issue, with a remarkably high mean of 0.96 (SD = 0.03) and a narrow interquartile range (0.95–0.98). This near-universal prevalence indicates that the vast majority of SDP datasets suffer from substantial feature irrelevance (Figure~\ref{boxplot_dimensionality}), consistent with prior empirical observations regarding the proliferation of weak predictive metrics in SDP datasets~\citep{jiarpakdee_study_2016}. The exceptionally low variance suggests this is a systemic characteristic of SDP datasets rather than dataset-specific anomaly, highlighting the critical need for feature selection methodologies in SDP practice.

\paragraph{\textit{Class Imbalance}} demonstrates high variability spanning the complete theoretical range (0.00–0.99) with a substantial mean of 0.67 (SD = 0.28). The wide interquartile range (0.55–0.91) indicates a bimodal distribution, where approximately 75\% of datasets exhibit moderate to severe imbalance (Figure~\ref{boxplot_imbalance}). This finding confirms class imbalance as both prevalent and highly variable, making it a primary concern for SDP model development and performance evaluation.

\paragraph{\textit{Class Overlap}} exhibits moderate prevalence with a mean of 0.21 (SD = 0.14), indicating that boundary ambiguity between defective and non-defective modules is common but varies substantially across datasets. The 75th percentile value of 0.31 suggests that approximately one-quarter of datasets experience significant class overlap (Figure~\ref{boxplot_overlap}), which has important implications for classifier performance and decision boundary complexity.

\paragraph{\textit{Outlier}} prevalence across the 374 datasets is relatively low in absolute proportion but shows noticeable variation (min = 0.02, max = 0.14, mean = 0.08, SD = 0.02). The narrow interquartile range (0.07–0.09) indicates that most datasets contain between 7\% and 9\% outlier values relative to their total feature space (Figure~\ref{boxplot_outlier}), suggesting a moderate but non-negligible degree of irregularity. These results imply that while extreme contamination is rare, a baseline level of outlier presence is the norm for SDP datasets.

\paragraph{\textit{Attribute Noise}} levels are also generally low across the dataset pool, with a mean of 0.05, and a right-skewed distribution (IQR: 0.01–0.05). This distribution (Figure~\ref{boxplot_noise}) indicates that the majority of datasets exhibit minimal non-informative variation in their attributes, but a small subset reaches noticeably higher noise levels.

\paragraph{\textit{\textbf{Co-occurrence of Data Quality Issues:}}} 
To complement the individual prevalence statistics, we examined how often different data quality issues appear together in the same dataset. Figure~\ref{co_occurrence_heatmap} presents a heatmap of pairwise co-occurrence counts across all 374 datasets. The results show that quality issues in SDP data almost never occur in isolation. For example, irrelevant features co-occur with outliers in every dataset and with class overlap in 372 datasets, reinforcing their role as a near-universal background condition. Similarly, class imbalance and class overlap co-occur in 368 datasets, suggesting that skewed class distributions and overlapping feature spaces frequently compound one another. Attribute noise shows the lowest pairwise co-occurrence (350–352 datasets), yet even this still represents more than 93\% of the dataset pool. Taken together, these patterns indicate that SDP datasets consistently exhibit multiple, interacting quality issues. This reinforces the need for joint rather than isolated analyses, as focusing on a single issue risks overlooking the conditions under which it co-occurs and interacts with others.

\begin{tcolorbox}[colback=gray!10,colframe=black,title=Answer to RQ1]

Our analysis of 374 datasets shows that data quality issues rarely occur in isolation. 
Irrelevant features are essentially universal (mean 0.96), class imbalance is highly prevalent 
(75\% of datasets exhibit moderate to severe imbalance), and class overlap is present in about 
one-quarter of datasets at significant levels. Outliers occur at a moderate baseline level (7--9\%), 
while attribute noise is the least frequent but still co-occurs with other issues in over 93\% of datasets. 
Thus, real-world SDP datasets must be understood as \emph{multi-problem environments} rather than 
single-issue cases.
\end{tcolorbox}

\subsection{Baseline Performance}
\label{sub2}
Next, to answer RQ2, we establish how widely used classifiers perform under these real-world conditions, without tuning, to reflect practical baseline usage. Although more elaborate validation schemes such as repeated cross-validation could have been employed, our aim here is not to optimize predictive accuracy but to characterize how co-occurring data quality issues condition baseline classifier behavior. A consistent single stratified train–test split provides a controlled and comparable foundation for our explanatory analyses. More complex resampling would primarily reduce estimate variance without altering the relative influence patterns that are the focus of this study. As such, the baseline performance results provide a reference point before analyzing how individual quality issues affect performance. Table~\ref{performance_stats} summarizes the balanced accuracy of the five selected classifiers across the 374 datasets.

\begin{table}[htbp]
\centering
\caption{Classifier Performance Summary}
\begin{tabular}{lcccccc}
\toprule
\textbf{Metric} & \textbf{DT} & \textbf{RF} & \textbf{NB} & \textbf{MLP} & \textbf{SVM} \\
\toprule
Min    & 0.43 & 0.45 & 0.28 & 0.39 & 0.48 \\
\midrule
Max    & 1.00 & 1.00 & 1.00 & 1.00 & 0.82 \\
\midrule
Mean   & 0.67 & 0.65 & 0.63 & 0.61 & 0.53 \\
\midrule
S.D.   & 0.15 & 0.15 & 0.12 & 0.12 & 0.06 \\
\midrule
25\%   & 0.57 & 0.52 & 0.56 & 0.50 & 0.50 \\
\midrule
75\%   & 0.73 & 0.73 & 0.70 & 0.69 & 0.52 \\
\bottomrule
\end{tabular}
\label{performance_stats}
\end{table}

Tree-based models show slightly higher average performance, with DT achieving the highest mean balanced accuracy (0.67, SD = 0.15), followed closely by RF (0.65, SD = 0.15). However, the relatively high variance suggests these models are sensitive to dataset-specific characteristics and may not generalize consistently across datasets with differing quality profiles.

Probabilistic (NB; 0.63, SD = 0.12) and neural (MLP; 0.61, SD = 0.12) models exhibit slightly lower but more consistent performance. Their lower variance suggests more stable behavior across the range of data quality conditions encountered, albeit with a trade-off in peak performance.

SVM shows the lowest average performance (0.53) but also the smallest variance (SD = 0.06). This pattern suggests not robustness, but rather a consistent underperformance. While this may in part reflect systematic sensitivity to prevalent data quality challenges in SDP datasets, it is also likely influenced by the use of default hyperparameter settings—SVMs are known to benefit substantially from kernel and parameter tuning. Unlike the other models, SVM appears less affected by dataset-specific variance and more constrained by a combination of baseline configuration and domain-specific challenges.

\begin{tcolorbox}[colback=gray!10,colframe=black,title=Answer to RQ2]

Decision Trees (mean = 0.67) and Random Forests (0.65) achieve the highest balanced accuracy, 
but their performance is highly variable, reflecting sensitivity to dataset-specific conditions. 
Naive Bayes (0.63) and MLP (0.61) show slightly lower means but greater stability. 
SVM has the lowest mean performance (0.53) but also the lowest variance, indicating 
consistent underperformance rather than robustness. 
Overall, there is a modest \emph{performance–robustness trade-off}: 
models with higher average performance also exhibit greater volatility across datasets.
\end{tcolorbox}

\subsection{Impact of Data Quality Issues}
\label{sub3}

\begin{figure}[htbp]
    \centering
    \includegraphics[width=0.9\textwidth]{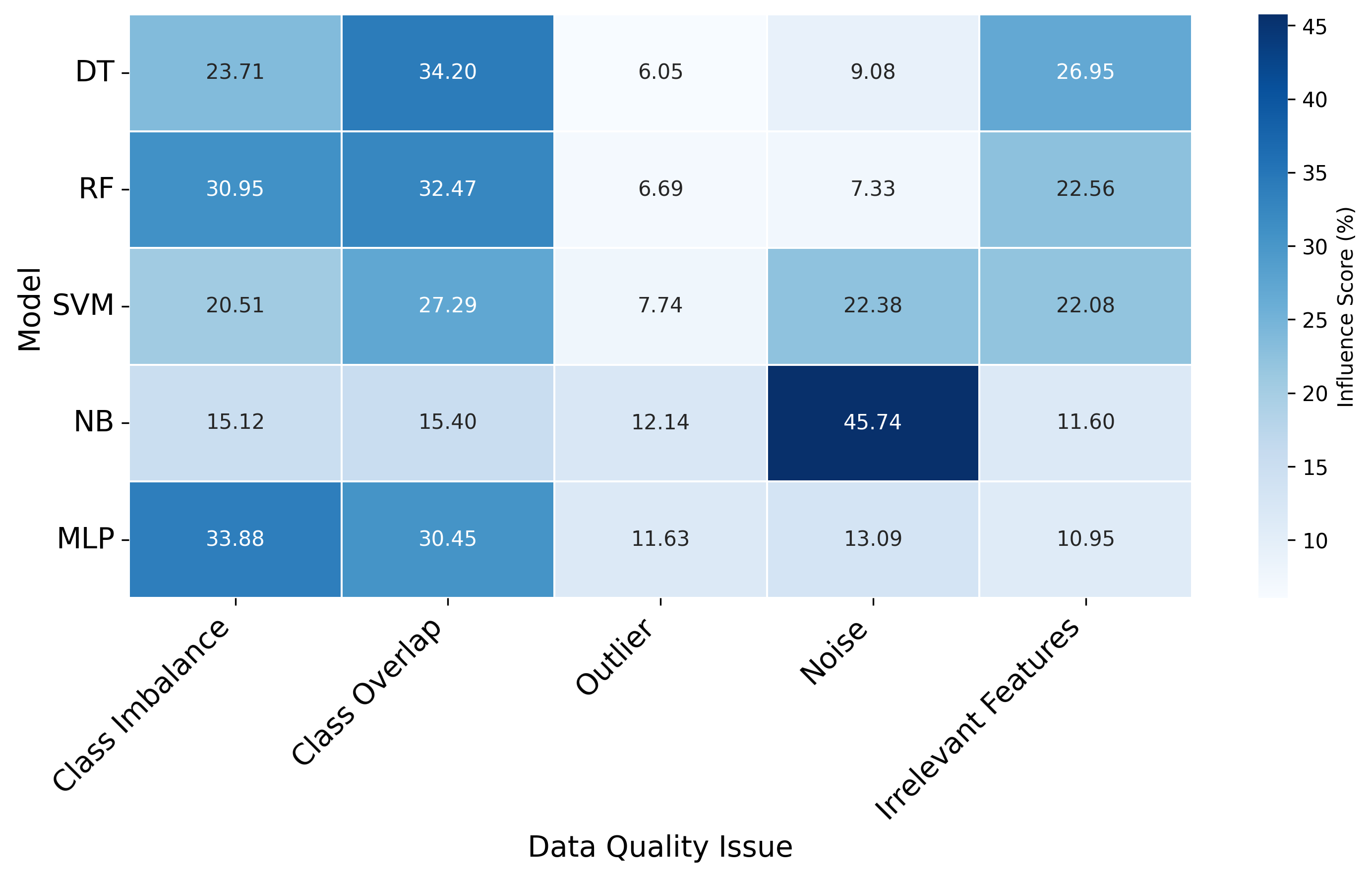}
    \caption{Heatmap illustrating the influence scores of data quality issues across different models. Darker colors indicate higher influence}
    \label{influence_heatmap}
\end{figure}

\begin{figure}[htbp]
    \centering
    \captionsetup{justification=centering} 
    \begin{subfigure}[b]{0.49\textwidth} 
        \centering
        \includegraphics[width=\textwidth]{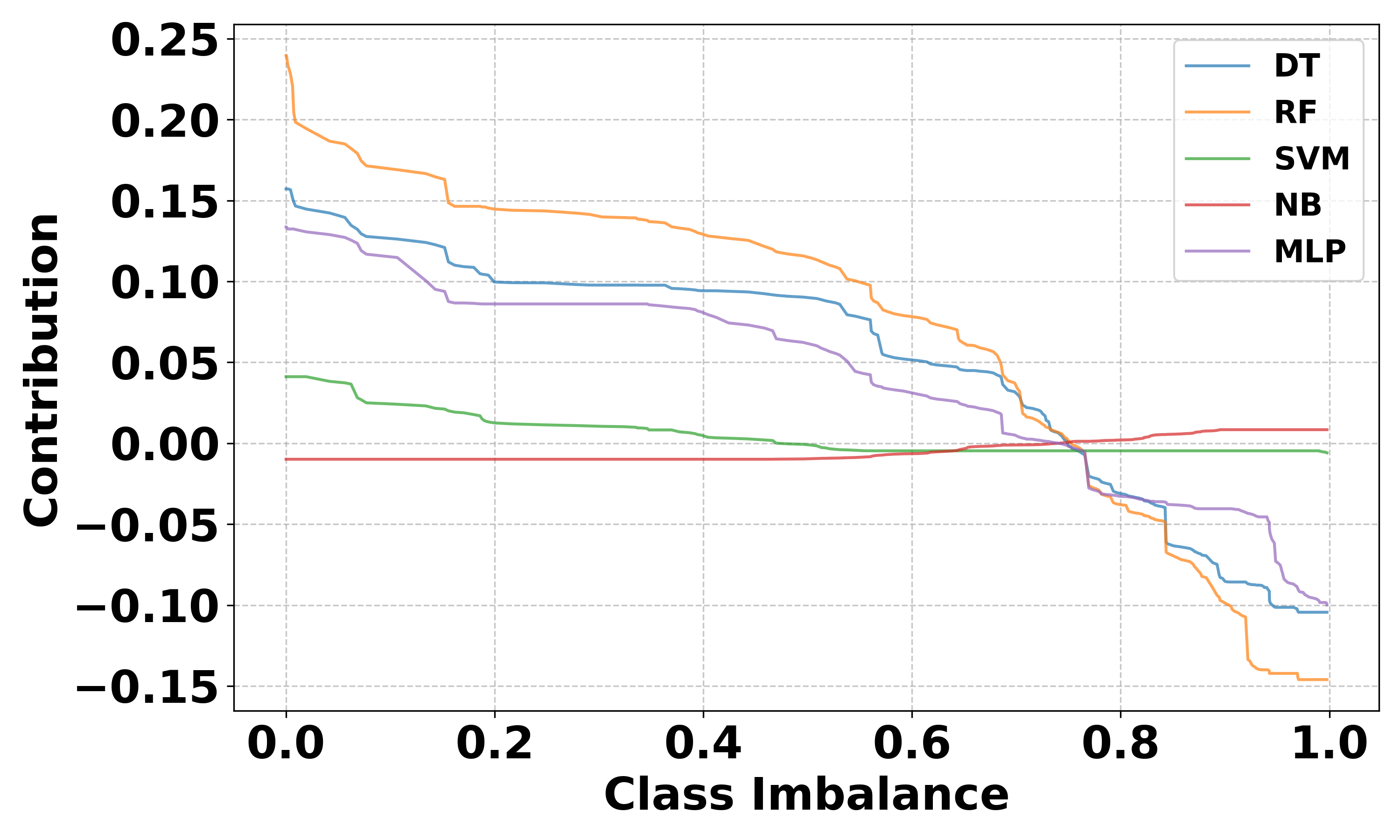} 
        \caption{Class Imbalance}
        \label{monotonic_c2}
    \end{subfigure}
    \hfill 
    \begin{subfigure}[b]{0.49\textwidth} 
        \centering
        \includegraphics[width=\textwidth]{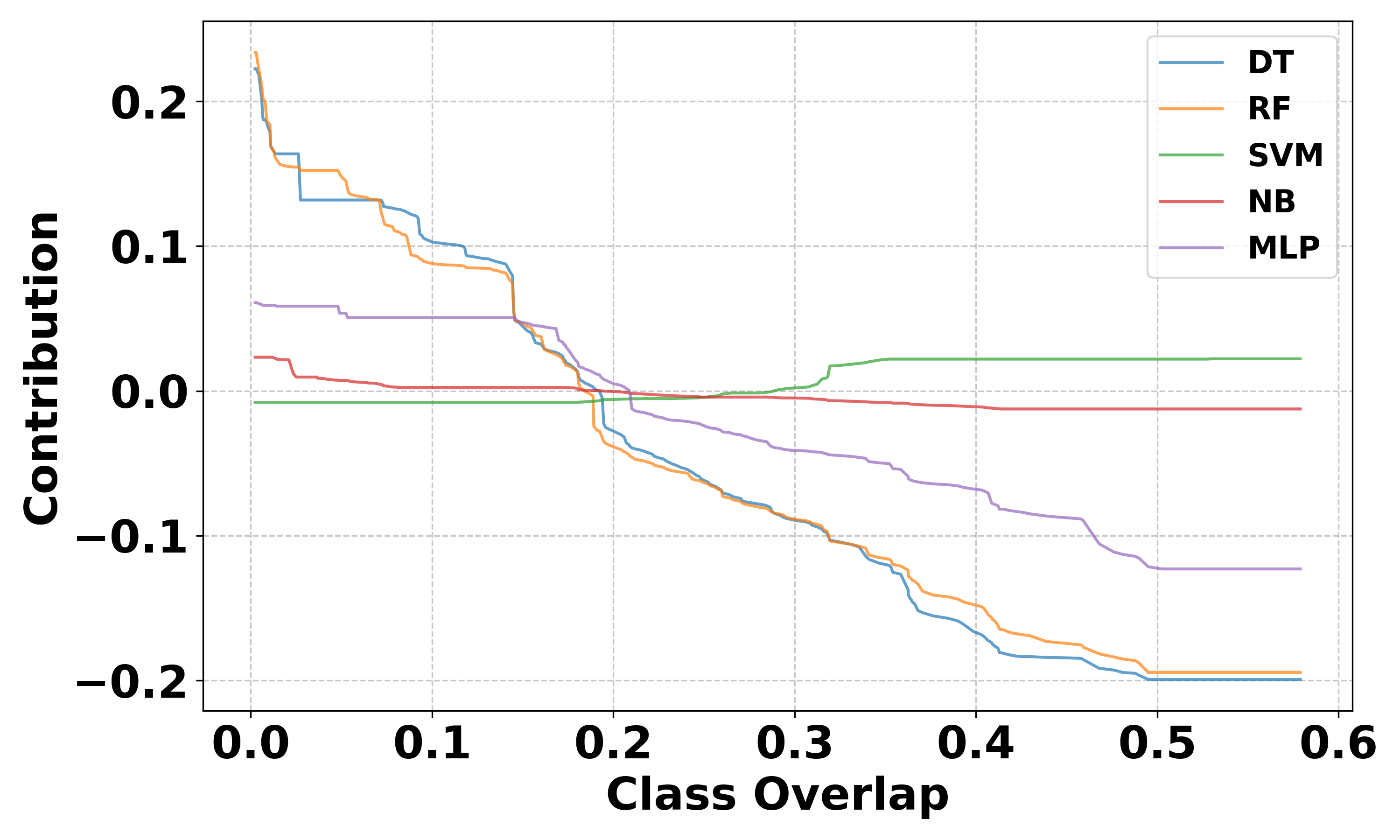} 
        \caption{Class Overlap}
        \label{monotonic_n1}
    \end{subfigure}
    \vspace{1em} 
    \begin{subfigure}[b]{0.49\textwidth} 
        \centering
        \includegraphics[width=\textwidth]{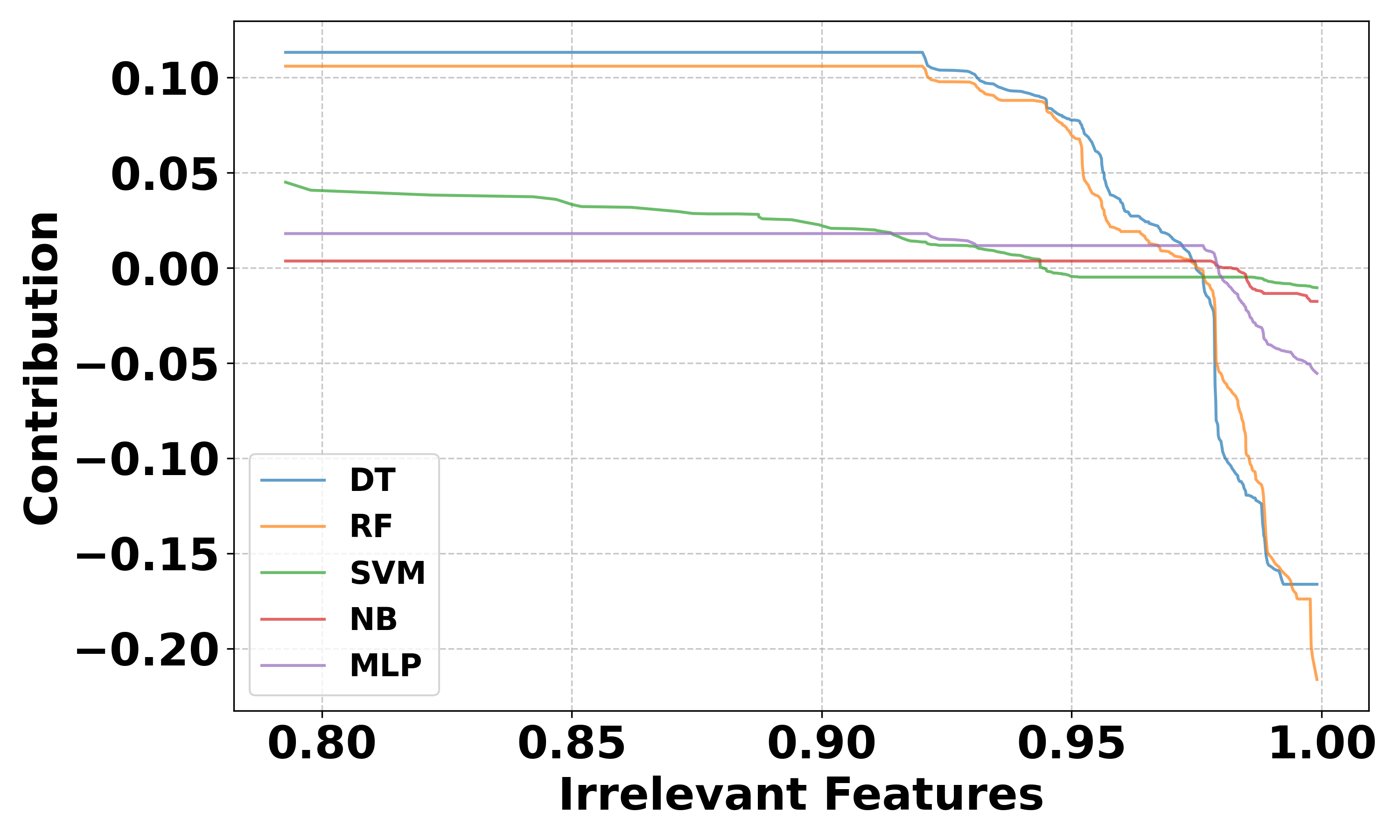} 
        \caption{Irrelevant Features}
        \label{monotonic_mut_inf}
    \end{subfigure}
    \hfill 
    \begin{subfigure}[b]{0.49\textwidth} 
        \centering
        \includegraphics[width=\textwidth]{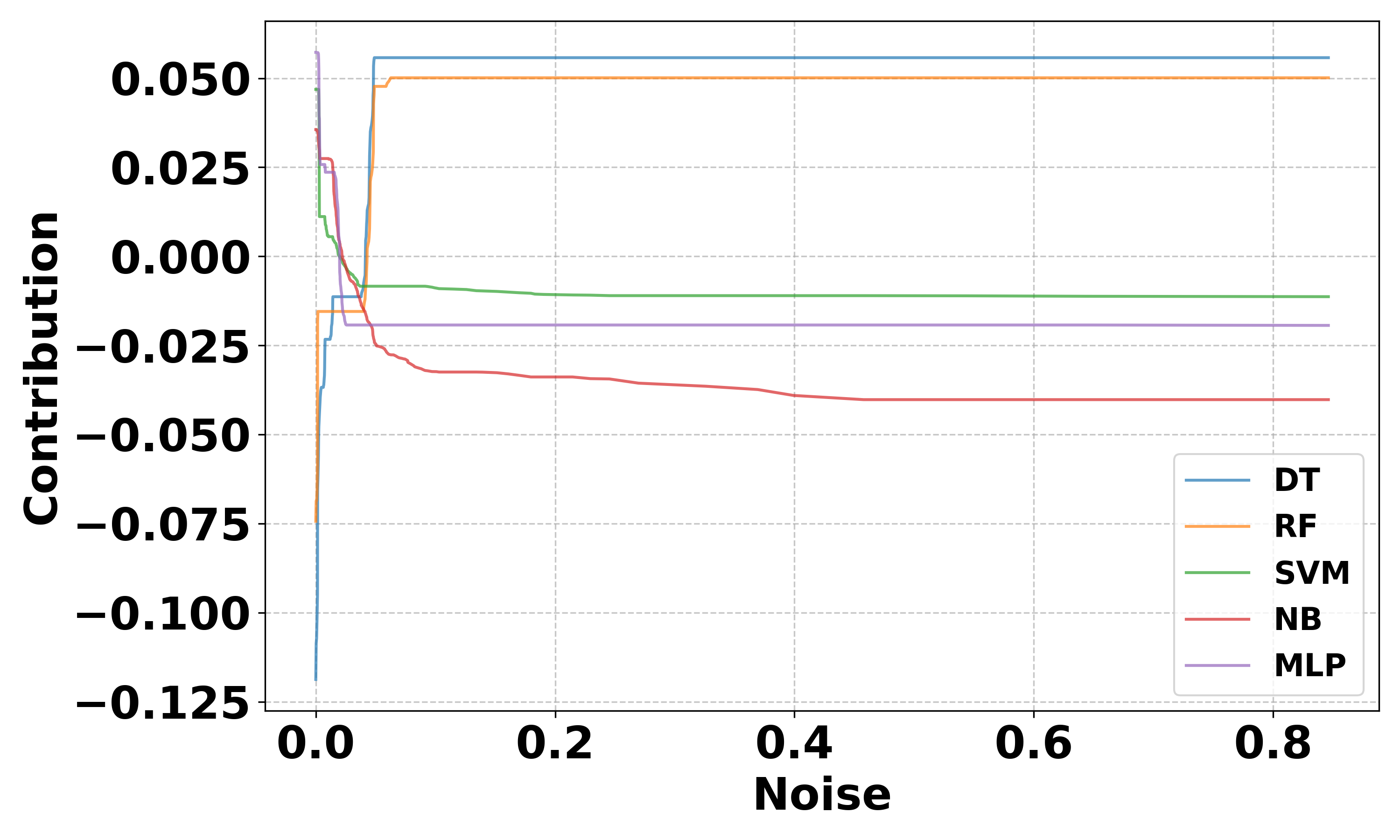} 
        \caption{Attribute Noise}
        \label{monotonic_noise}
    \end{subfigure}
    \vspace{1em} 
    \begin{subfigure}[b]{0.49\textwidth} 
        \centering
        \includegraphics[width=\textwidth]{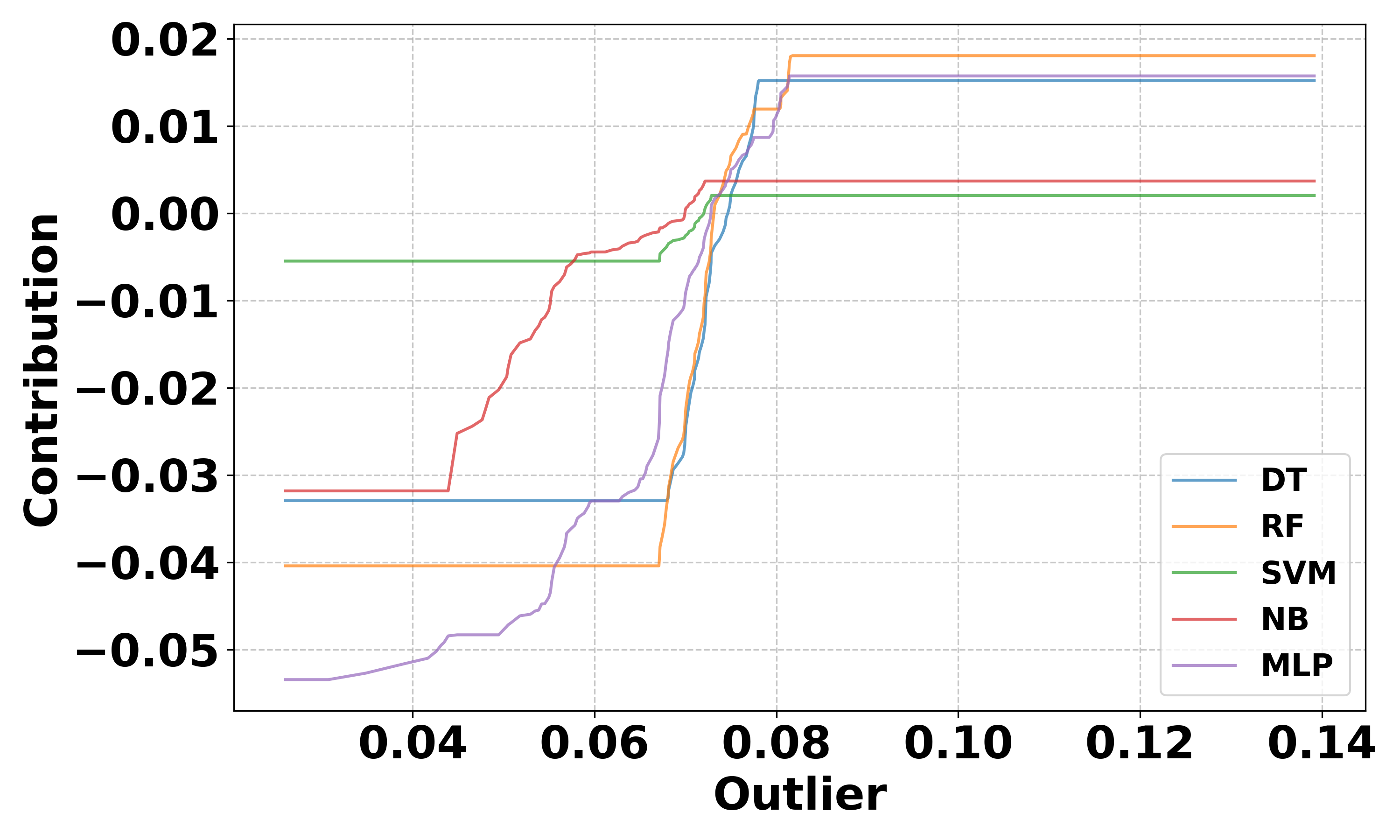} 
        \caption{Outlier}
        \label{monotonic_outliers}
    \end{subfigure}

    \caption{Monotonic relationship between each data quality issue and model performance. The x-axis shows the severity of the issue, and the y-axis represents its additive contribution to Balanced Accuracy relative to the EBM intercept} 
    \label{monotonic_relationships} 
\end{figure}

\begin{figure}[htbp]
    \centering
    \captionsetup{justification=centering} 
    \begin{subfigure}[b]{0.49\textwidth} 
        \centering
        \includegraphics[width=\textwidth]{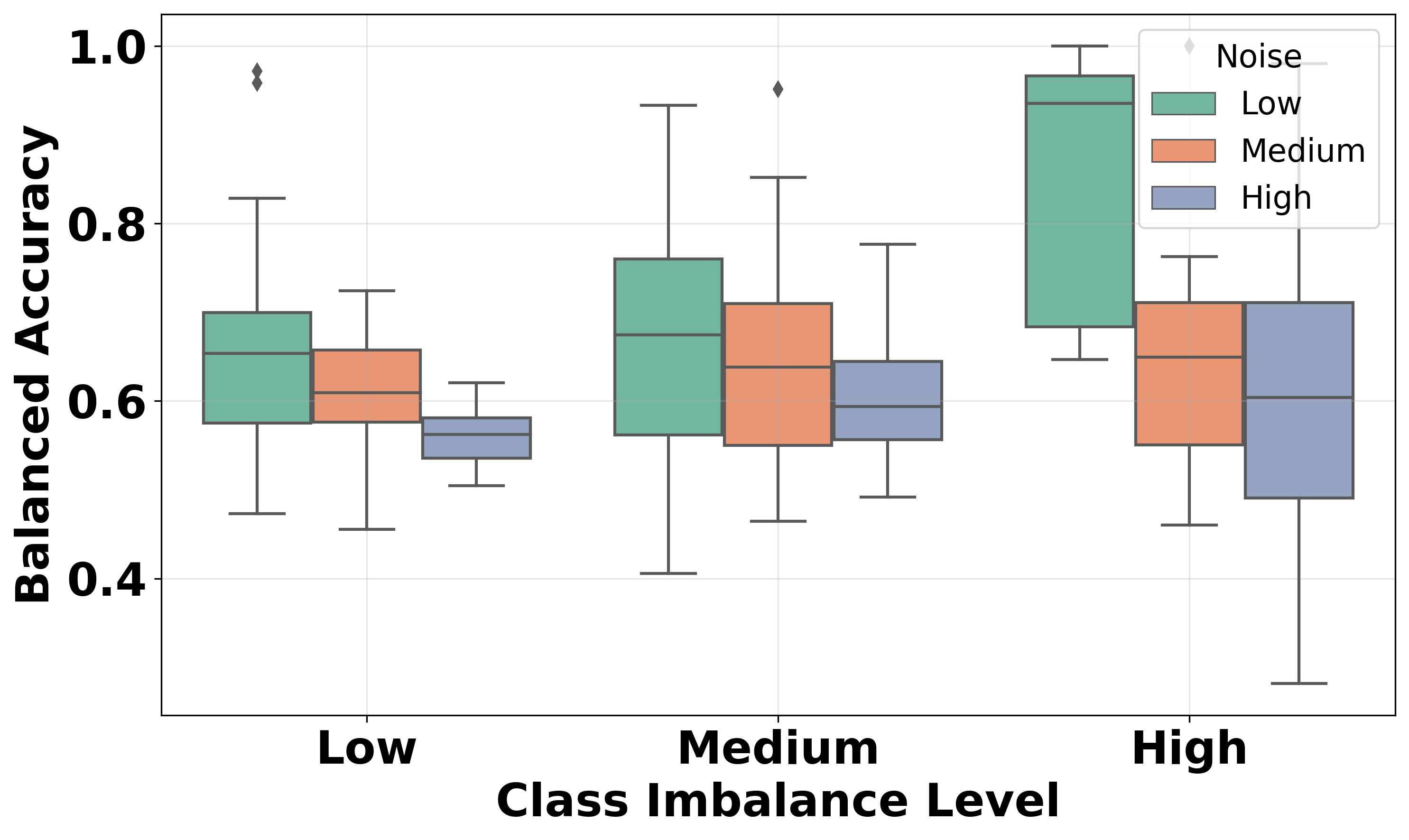} 
        \caption{Class Imbalance, Stratified by Noise (NB)}
        \label{c2_vs_noise_NB}
    \end{subfigure}
    \hfill 
    \begin{subfigure}[b]{0.49\textwidth} 
        \centering
        \includegraphics[width=\textwidth]{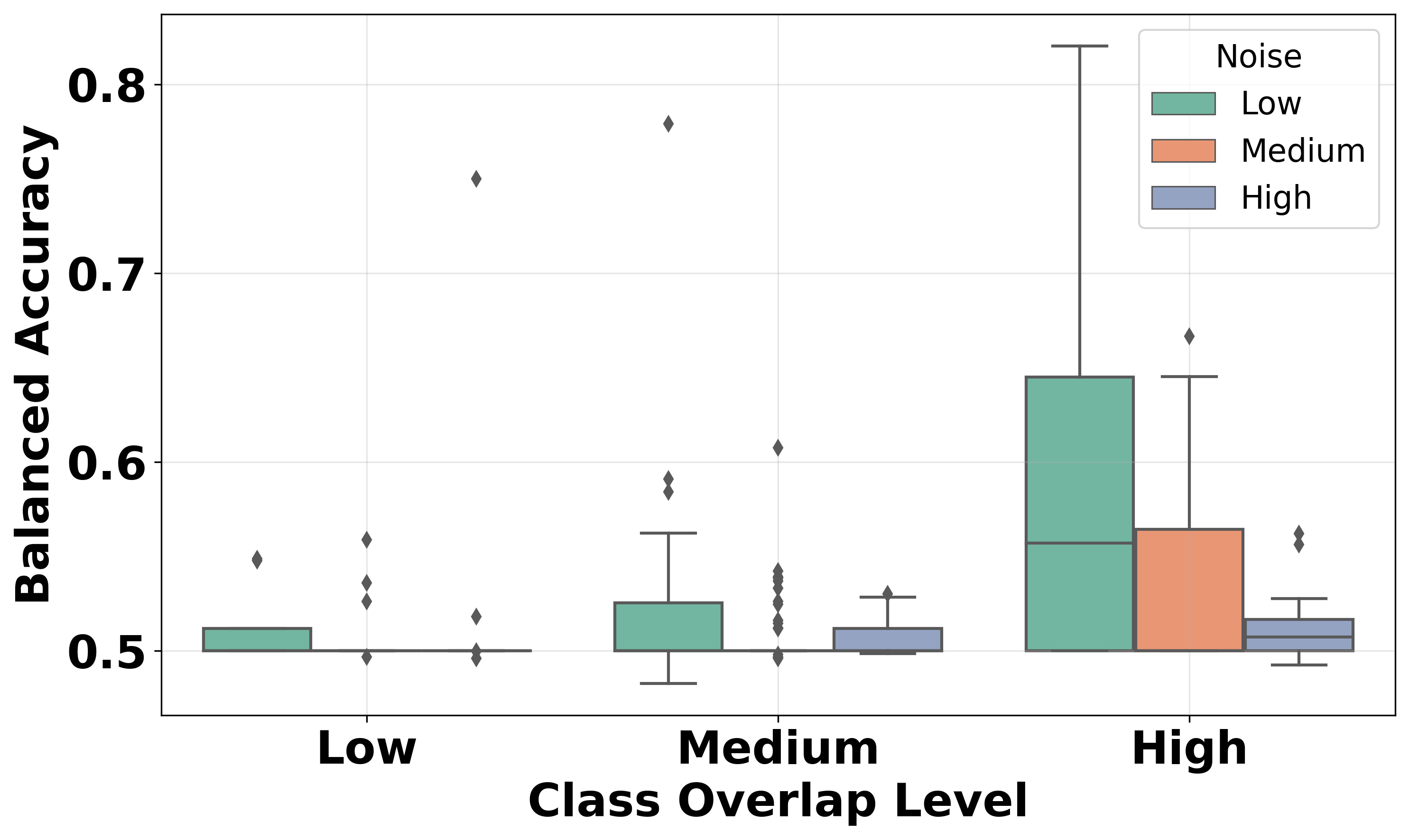} 
        \caption{Class Overlap, Stratified by Noise (SVM)}
        \label{n1_vs_noise_SVM}
    \end{subfigure}    
    \vspace{1em} 
    
    \begin{subfigure}[b]{0.49\textwidth} 
        \centering
        \includegraphics[width=\textwidth]{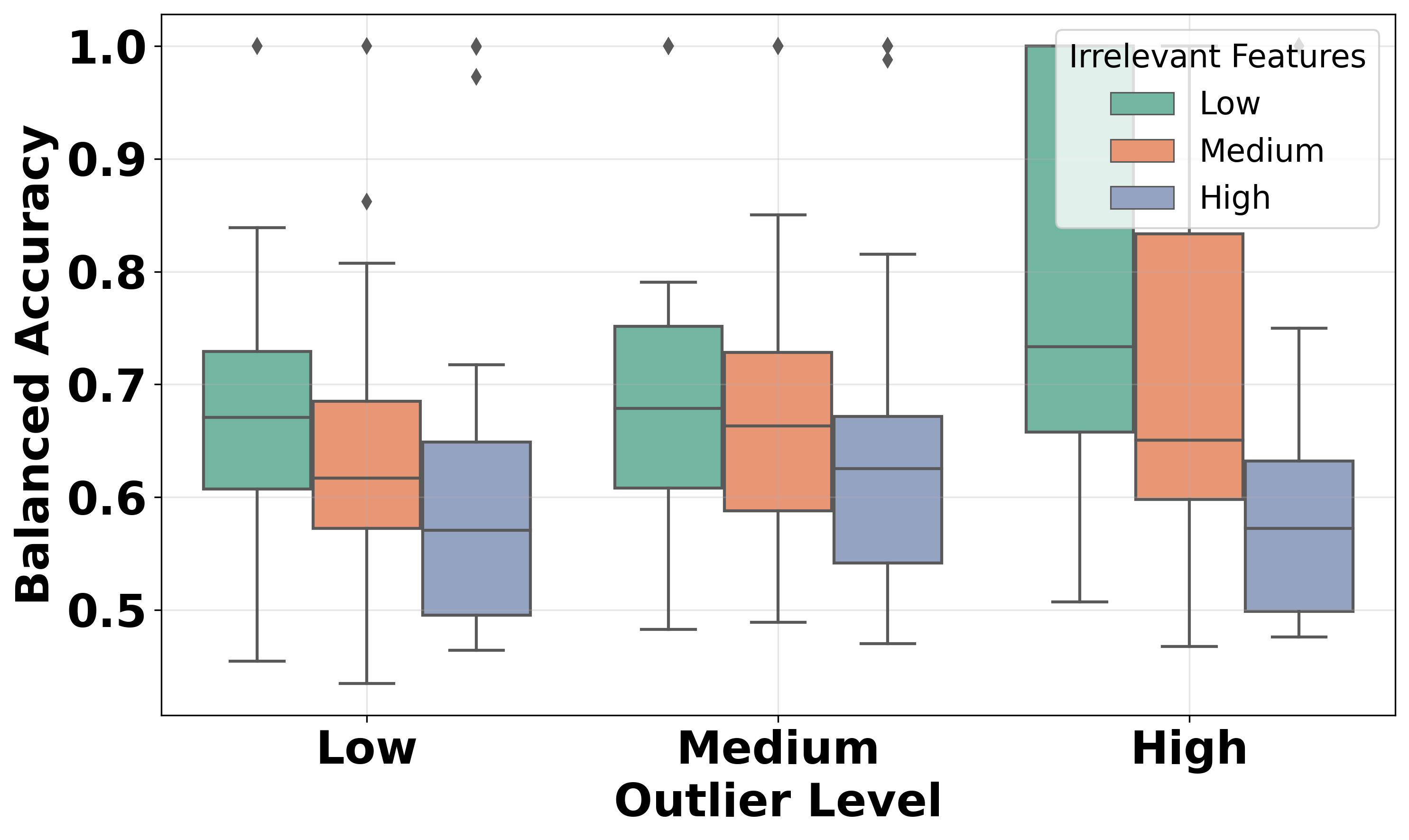} 
        \caption{Outlier, Stratified by Irrelevant Features (DT)}
        \label{outlier_vs_irrelevant_features_DT}
    \end{subfigure}
    \hfill 
    \begin{subfigure}[b]{0.49\textwidth} 
        \centering
        \includegraphics[width=\textwidth]{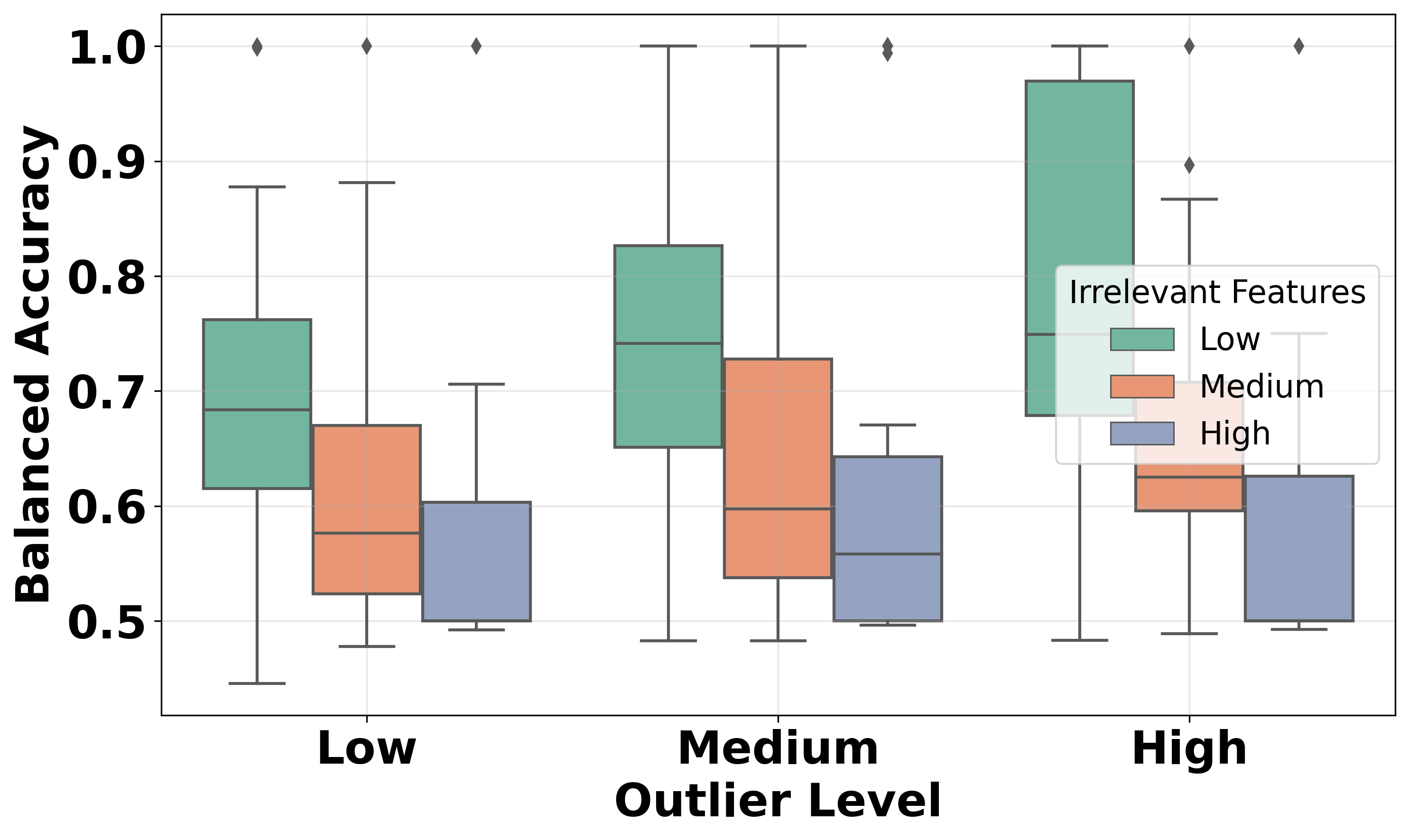} 
        \caption{Outlier, Stratified by Irrelevant Features (RF)}
        \label{outlier_vs_irrelevant_features_RF}
    \end{subfigure}
    \vspace{1em} 
    \begin{subfigure}[b]{0.49\textwidth} 
        \centering
        \includegraphics[width=\textwidth]{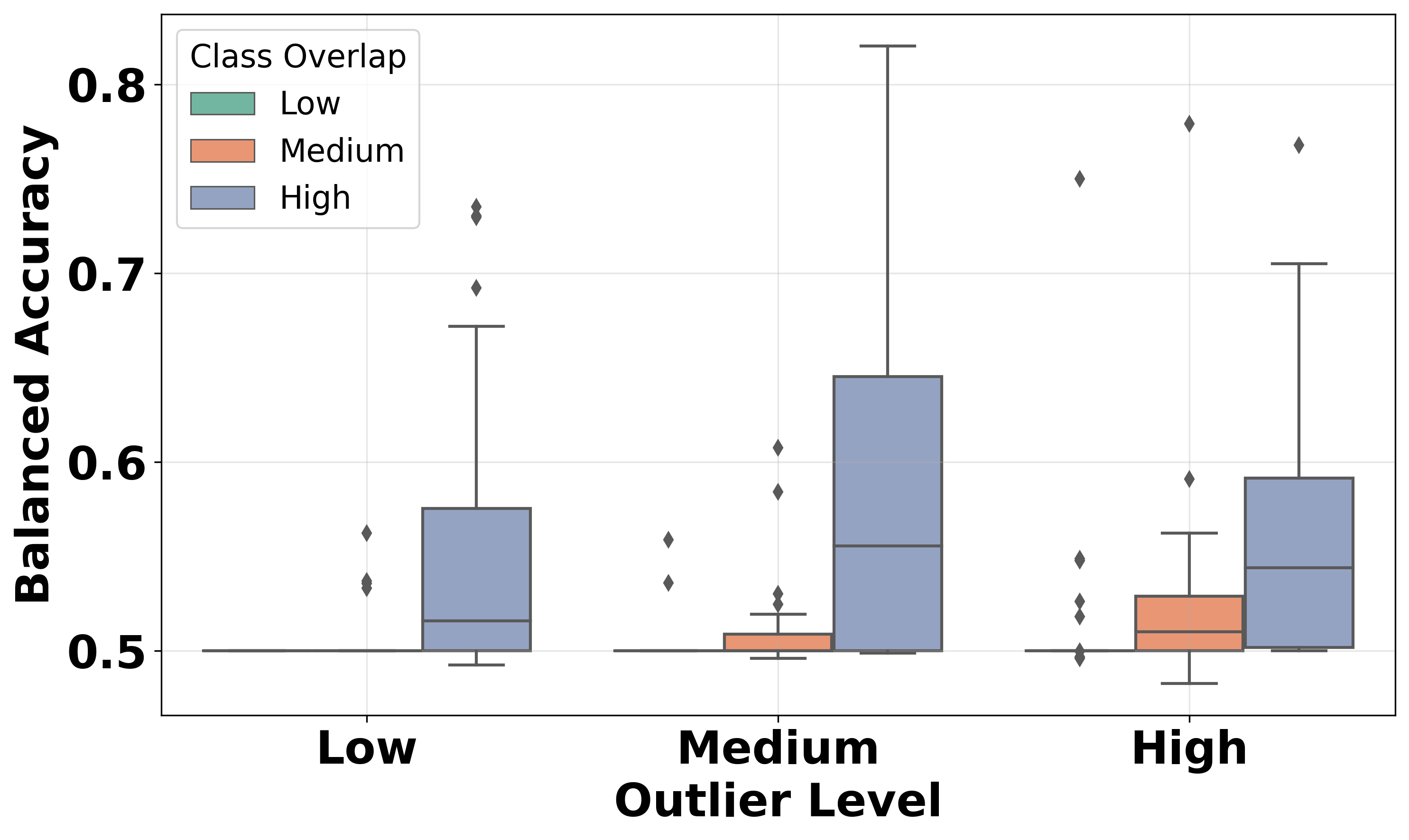} 
        \caption{Outlier, Stratified by Class Overlap (SVM)}
        \label{outlier_vs_n1_SVM}
    \end{subfigure}
    \hfill 
    \begin{subfigure}[b]{0.49\textwidth} 
        \centering
        \includegraphics[width=\textwidth]{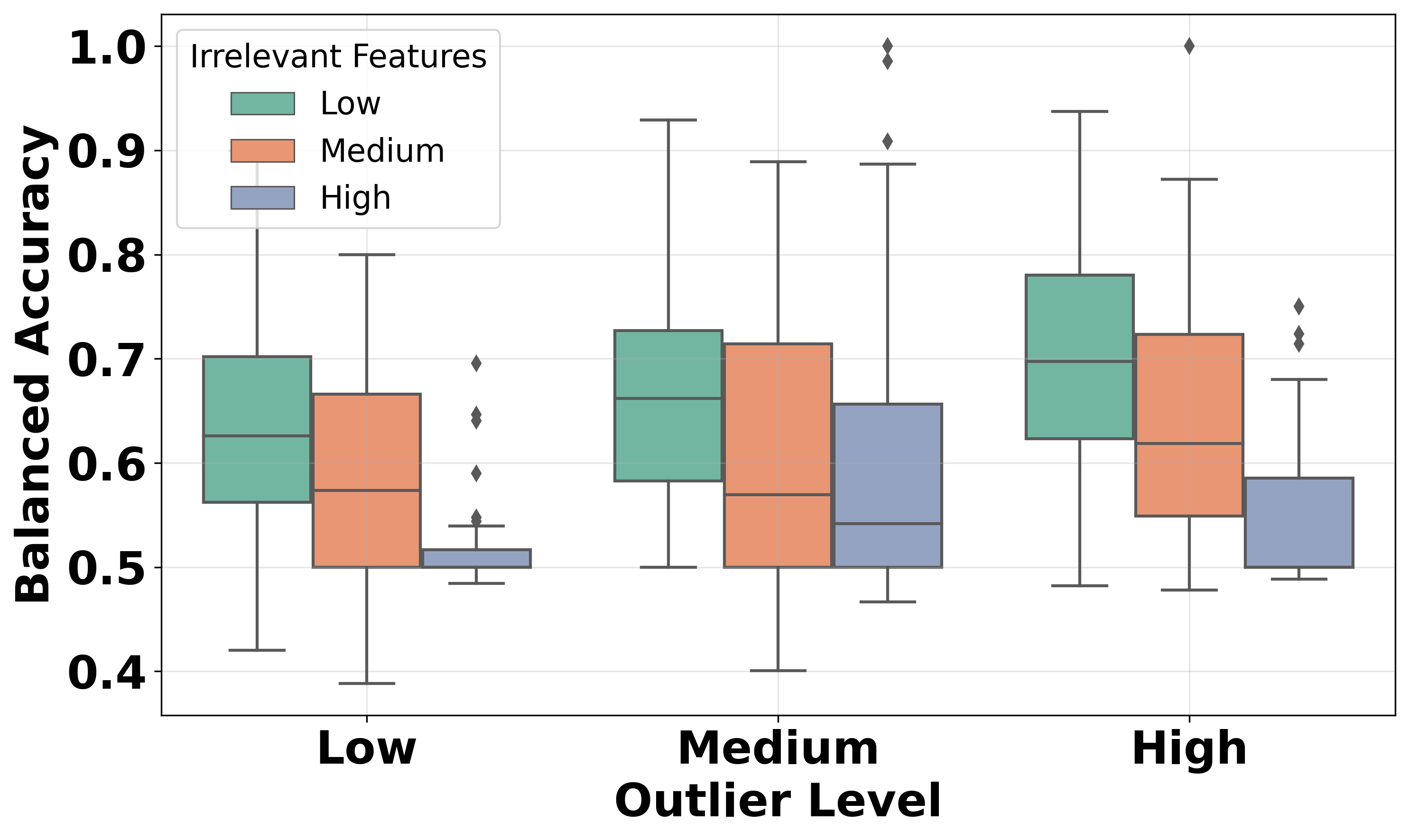} 
        \caption{Outlier, Stratified by Irrelevant Features (MLP)}
        \label{outlier_vs_irrelevant_features_MLP}
    \end{subfigure}
    
    \caption{Stratified Interaction Analysis with Boxplots} 
    \label{stratified_analysis} 
\end{figure}

Finally, RQ3 examines the relative influence of each quality issue on model performance, including thresholds where effects shift and interactions that reveal conditional or counterintuitive patterns. Figure~\ref{influence_heatmap} shows the influence heatmap, Figure~\ref{monotonic_relationships} depicts the shape functions and Figure~\ref{stratified_analysis} illustrates the interaction effect.

\paragraph{\textbf{Class Imbalance}} emerges as one of the most impactful data quality issues across models (Figure~\ref{influence_heatmap}). Its influence is strongest for MLP (33.88\%), RF (30.95\%), and DT (23.71\%), while SVM (20.51\%) and NB (15.12\%) show comparatively lower sensitivity. This pattern suggests that the performance of tree-based and neural models is more strongly conditioned on class distribution, whereas SVM and NB are somewhat less dependent on balance between classes.

The EBM shape functions (Figure~\ref{monotonic_c2}) further clarify these effects. For DT, RF, and MLP, balanced or near-balanced data ($\leq 0.2$ imbalance) contributes positively to performance, indicating that these models benefit from the availability of sufficient examples from both classes. As imbalance increases, however, their performance declines steadily, with sharp negative contributions beyond ~0.7 imbalance. Notably, RF—despite its ensemble robustness—shows the steepest drop under severe imbalance.

A consistent tipping point emerges across nearly all models: at imbalance values of approximately 0.65–0.70, the effect on balanced accuracy shifts from positive to negative. This threshold appears for tree-based, neural, and even SVM models, suggesting a general statistical boundary beyond which the minority class becomes too sparse for reliable decision boundary learning. While the magnitude of the decline varies—RF dropping most sharply, SVM more gradually—the location of the turning point is strikingly stable.

For NB, imbalance exerts minimal influence overall, with the contribution curve remaining close to zero across the full spectrum. An apparent slight positive drift at extreme imbalance contradicts prior studies~\citep{eldho_2022,liu_comparative_2022}. Stratified analysis (Figure~\ref{c2_vs_noise_NB}) shows that this pattern is not a true benefit of imbalance, but rather a co-occurrence artefact: when imbalance is high but other harmful issues, such as attribute noise, are low, NB can maintain or slightly improve balanced accuracy. This favorable combination produces a small net-positive effect in the aggregated EBM curve, though its magnitude is negligible in practice.

Taken together, these results highlight two key insights. First, class imbalance is a major driver of performance degradation for most models, with a clear and consistent tipping point around 0.65–0.70. Second, interpreting imbalance effects without accounting for co-occurring issues can be misleading, as interactions may mask or even reverse apparent trends.

\paragraph{\textbf{Class Overlap}} emerges as an even more immediate source of degradation. It is the most influential data quality issue for DT (34.20\%), RF (32.47\%), and MLP (30.45\%), while its effect is moderate for SVM (27.29\%) and relatively low for NB (15.40\%). This pattern suggests that tree-based and neural learners are especially vulnerable when classes intermingle in feature space, whereas SVM and NB are comparatively less sensitive.

The EBM shape functions (Figure~\ref{monotonic_n1}) reveal a clear non-linear trend for the sensitive models. DT and RF show almost identical curves: strong positive contributions at very low overlap ($\leq 0.1$), followed by a steady decline that crosses into negative territory at approximately 0.18–0.20. Beyond this point, performance deteriorates sharply, reaching its lowest values once overlap exceeds 0.4. MLP follows the same general trajectory, but with a smaller initial gain and an earlier onset of decline. Its contributions then flatten into a negative plateau (just below –0.1) once overlap surpasses 0.35–0.40, indicating that additional overlap beyond this range no longer worsens performance substantially.

A notable contrast with class imbalance is the location of the tipping point: for overlap, degradation begins around ~0.20, far earlier than the ~0.65–0.70 threshold observed for imbalance. This shows that overlap is a more immediate and unavoidable challenge, capable of harming performance even at moderate levels. For SVM, the curve remains largely flat, with a slight upward drift beyond 0.25. Stratified interaction analysis (Figure~\ref{n1_vs_noise_SVM}) reveals this as a co-occurrence artefact: datasets with high overlap and concurrently low attribute noise produce stronger results, obscuring the generally negative effect in the aggregated model. NB remains close to zero across the full range, reflecting limited sensitivity—likely because its probabilistic independence assumptions make it less reliant on sharply separated class boundaries.

In sum, overlap exerts its influence at lower thresholds than imbalance, underscoring its role as a particularly damaging issue for tree-based and neural models. At the same time, the interaction artefacts observed for SVM highlight why overlap, like imbalance, must be interpreted in the context of co-occurring conditions.

\paragraph{\textbf{Irrelevant Features}} appear as a secondary but still non-negligible source of performance degradation. The influence score heatmap (Figure~\ref{influence_heatmap}) shows that DT (26.95\%) and RF (22.56\%) are most sensitive, with SVM (22.08\%) close behind. NB (11.60\%) and MLP (10.95\%) record the lowest influence, indicating that in practice their performance is less dependent on irrelevant feature removal compared to issues like imbalance or overlap.

The EBM shape functions (Figure~\ref{monotonic_mut_inf}) highlight the strongly threshold-driven nature of this effect. Across all models, contributions remain stable and slightly positive until the proportion of irrelevant features approaches ~0.94. This stability indicates that moderate levels of irrelevant features do not substantially hinder learning. However, once this tipping point is crossed, performance declines, with the steepness of the drop varying by model family.

DT and RF show the most pronounced vulnerability: they hold steady at ~+0.11 contribution up to the threshold, then fall sharply, with RF dropping below –0.15 at maximum irrelevance. This aligns with their greedy feature-splitting mechanisms—when irrelevant features dominate, the risk of uninformative splits rises dramatically, compounding errors within the tree or ensemble. SVM begins its decline slightly earlier than tree-based models, indicating that it is more sensitive to moderate levels of irrelevant features. However, the decline is more gradual, suggesting that while SVM is less tolerant at the onset, it degrades more slowly as irrelevance becomes extreme. In practice, this means SVM performance starts to erode sooner but avoids the sharp collapses observed for DT and RF. MLP maintains near-flat, slightly positive contributions until ~0.96, before showing a moderate decline, reflecting partial tolerance but eventual susceptibility once informative features are vastly outnumbered. NB remains largely unaffected, staying near zero until ~0.98 irrelevance, with only a slight negative drift thereafter.

A striking result is the stability of the threshold: across all models, the transition from neutral/positive to negative occurs consistently around 0.94. While the severity of the drop depends on the learning algorithm, the location of this tipping point echoes the consistent thresholds already observed for imbalance (~0.65–0.70) and overlap (~0.20). In sum, irrelevant features may not always be the most dominant factor, but they impose a sharp and predictable performance penalty once they exceed a critical proportion. This reinforces the broader picture: different issues degrade models at different thresholds, but all exhibit stable transition points that practitioners can use as warning markers for intervention.

\paragraph{\textbf{Attribute Noise}} also emerges as a secondary or tertiary concern for most models, with one notable exception: NB (45.74\%), for which noise is the single most influential issue. SVM (22.38\%) and MLP (13.09\%) show moderate sensitivity, while DT (9.08\%) and RF (7.33\%) record the lowest relative influence.

The EBM shape functions (Figure~\ref{monotonic_noise}) reveal distinct patterns across model families. DT and RF show a small negative contribution at near-zero noise, followed by a shallow positive plateau (~+0.05) that persists across most of the range. This trajectory can be understood in terms of tree-based dynamics: minimal noise disrupts early splits, reducing informativeness and producing slight performance loss. Once this noisy background is established, however, additional small amounts of noise have little further impact, and ensemble averaging in RF smooths outcomes, creating the appearance of a flat or slightly positive plateau. In contrast, SVM, MLP, and especially NB remain consistently negative across the full noise spectrum, with NB showing the steepest initial decline before flattening out.

A common tipping point is visible across models at very low noise values (0.02–0.05), where contributions shift sharply before stabilizing into flat plateaus in different directions. Importantly, our stratified analyses with imbalance, overlap, irrelevance, and outliers revealed no systematic conditions under which noise reliably improves tree-based performance. The shallow plateau for DT and RF therefore reflects a leveling-off of harm rather than substantive evidence that noise is beneficial. Overall, attribute noise plays only a secondary role compared to imbalance, overlap, and irrelevance, but it remains a decisive factor for NB, where it dominates model behavior across the full range.

\paragraph{\textbf{Outliers}} play a comparatively minor role across most models, with influence scores ranging from 6.05\% (DT) to 12.14\% (NB), consistently lower than for class imbalance, class overlap, or attribute noise (Figure~\ref{influence_heatmap}).

The EBM shape functions (Figure~\ref{monotonic_outliers}) show that DT, RF, and MLP follow a similar trajectory: small negative contributions at low outlier levels, transitioning around the 0.07–0.08 threshold into a shallow positive plateau (+0.01 to +0.02). Stratified analyses clarify when these counterintuitive upticks occur. For DT and RF, improvements appear only when irrelevant features are low (Figures~\ref{outlier_vs_irrelevant_features_DT},~\ref{outlier_vs_irrelevant_features_RF}), suggesting that outliers can help sharpen decision boundaries when the feature space is relatively clean. SVM shows modest gains primarily in high class-overlap scenarios (Figure~\ref{outlier_vs_n1_SVM}), consistent with its margin-based learning: outliers can act as additional support vectors that refine complex boundaries. MLP exhibits small benefits in settings with low irrelevance (Figure~\ref{outlier_vs_irrelevant_features_MLP}), where outliers may act as extra informative cases that enrich its learned representations. NB shows a different pattern. Its apparent positive drift could not be traced to any consistent interaction with other quality issues. Instead, it is likely a consequence of its probabilistic formulation: outliers distort Gaussian class-conditional distributions in ways that can exaggerate separability between classes, creating the appearance of improved performance. These gains therefore reflect a quirk of the model’s independence assumptions rather than genuine robustness.

Overall, outliers exert only a minor global influence compared to other quality issues, but their local effects depend strongly on data conditions and model assumptions. For tree-based and margin-based learners, they can sometimes clarify boundaries; for NB, they produce spurious gains rooted in distributional artefacts. These results show that outliers are not uniformly harmful, but neither should they be assumed beneficial. This cautions against automatic removal and supports profiling their role in the broader context of co-occurring quality issues before deciding on treatment.

\paragraph{\textit{\textbf{Threshold Summary:}}}
Across the five data quality issues, our analyses reveal consistent tipping points where model performance shifts from neutral or positive to negative. Specifically, (i) \textbf{class overlap} begins to harm models at approximately \textbf{0.18--0.20}, (ii) \textbf{class imbalance} becomes detrimental beyond \textbf{0.65--0.70}, and (iii) \textbf{irrelevant features} trigger sharp declines once they exceed about \textbf{0.94} of the feature set. Attribute noise and outliers exhibit more model-specific or interaction-dependent effects, but the three thresholds above are strikingly stable across classifiers. These values provide practical cut-offs for anticipating when data quality issues are likely to degrade performance.

\begin{tcolorbox}[colback=gray!10,colframe=black,title=Answer to RQ3]

Our analysis shows that data quality issues differ in both their overall influence and the way they degrade model performance. Class overlap, imbalance, and irrelevance consistently exert the strongest impact, while attribute noise and outliers are secondary and highly context-dependent. For the dominant issues, performance declines follow stable patterns across models, whereas noise and outliers show model-specific or interaction-driven behaviors. These results highlight that issue effects cannot be interpreted in isolation: co-occurrence can mask, amplify, or even invert expected trends. No classifier is universally robust, underscoring the importance of dataset profiling and interaction-aware evaluation when selecting or configuring models.
\end{tcolorbox}

\section{Implications}
\label{implications}
Our findings carry important implications at three levels: for practitioners building SDP models, for researchers in empirical software engineering, and for the broader data-centric machine learning community.

\subsection{Implications for Practitioners}
\begin{itemize}
    \item The near-universal co-occurrence of data quality issues (even the least frequent, attribute noise, appears in 93\% of datasets) implies that real-world SDP data must be understood as multi-problem environments rather than single-issue cases. This affects how practitioners interpret model performance and motivates multi-faceted preprocessing. 
    
    \item The consistent tipping points identified—overlap around 0.20, imbalance around 0.65, and irrelevance around 0.94—imply that there are natural thresholds beyond which standard learners begin to degrade. These thresholds act as practical markers for anticipating when additional intervention or alternative learners may be needed.
    
    \item The observation that no model is universally robust but instead sensitive to specific issues implies that model choice in practice is conditional: DT and RF excel in balanced, relevant features settings but fail under extreme feature irrelevance or class imbalance, while SVM is resilient to overlap but underperforms under defaults. 
    
    \item The finding that outliers can in some contexts improve rather than harm performance implies that common preprocessing practices of automatic outlier removal may be counterproductive, especially in settings with low feature irrelevance or high overlap.
\end{itemize}

\subsection{Implications for Researchers}
\begin{itemize}
    \item The variability of effects across co-occurring issues implies that reporting only aggregate performance is insufficient. Research in SDP must account for dataset-specific quality profiles to enable fair cross-study comparisons. 
    
    \item The discovery of masked or inverted effects under stratified analysis implies that interaction effects are central to understanding data quality. This challenges the current norm of studying issues in isolation and highlights the need for multi-factor analyses. 
    
    \item The reliance on defaults implies that observed robustness patterns reflect baseline usage rather than performance ceilings. This underscores the importance of replication with tuned and cost-sensitive configurations in future work. 
    
    \item The trade-off between performance and robustness across models implies that future evaluation should include robustness metrics (variance, stability across strata), not just central accuracy measures.
\end{itemize}

\subsection{Broader Implications Beyond SDP}
Although grounded in defect prediction, the findings imply a broader data-centric principle: robustness depends less on selecting the ``best'' model globally and more on aligning model choice with dataset-specific quality profiles. For empirical software engineering, this reframes defect prediction research as a data-aware problem; for machine learning at large, it illustrates that quality profiling should be treated as a first-class component of modeling pipelines.

\section{Threats to Validity}
\label{threats_to_validity}
While this study offers important insights into the relative effects of data quality issues on SDP models, several threats to validity should be considered when interpreting the results.

\paragraph{Internal Validity.}

\begin{itemize}
    \item \textbf{Evaluation Protocol:} Our use of a single stratified train–test split, rather than repeated cross-validation, may increase variance in individual performance estimates. However, since the study’s goal is to characterize relative influence patterns of data quality issues rather than maximize predictive accuracy, we expect the overall conclusions to remain robust.
    \item \textbf{Default Hyperparameters:} All classifiers were evaluated using their default configurations in \texttt{scikit-learn}~\citep{scikit-learn}. While this ensures consistency and reflects common usage in practice, it may underestimate performance under tuned settings. However, this design choice avoids bias introduced by selective tuning and ensures fair cross-model comparison.
    
    \item \textbf{Influence Score Estimation:} The use of Explainable Boosting Machines assumes an additive contribution of quality metrics. While EBMs can capture non-linearities and some interactions, they may not fully model higher-order dependencies. We partially mitigate this by supplementing EBMs with stratified interaction analysis.
    
    \item \textbf{Dataset Processing Choices:} Preprocessing steps (e.g., imputation, binary label harmonization) could introduce artefacts. However, we applied consistent procedures across all datasets.
\end{itemize}

\paragraph{Construct Validity.}

\begin{itemize}
    \item\textbf{Metric Definitions:} Our measures for data quality issues are grounded in prior work~\citep{lorena2020,michie_1995}, but alternative formulations exist (e.g., different overlap or outlier detection criteria). We selected metrics that are both interpretable and widely accepted, though future work could examine robustness to alternative formulations.
\end{itemize}

\paragraph{External Validity.}

\begin{itemize}
    \item \textbf{Dataset Representativeness:} The experimental dataset pool comprises 374 SDP datasets, primarily collected from established benchmark repositories. While this ensures relevance and accessibility, the dataset characteristics may not generalize to all real-world SDP tasks, especially those from industrial contexts.
    
    \item \textbf{Model Scope:} The study evaluates five well-known classifiers: DT, RF, NB, SVM, and MLP. Although these represent diverse learning paradigms, they do not cover emerging or specialized models such as deep learning architectures, graph-based learners, or recent ensembles. The observed effects may not extrapolate to all model families.
\end{itemize}

\paragraph{Conclusion Validity.}

\begin{itemize}
    \item\textbf{EBM Interpretability vs. Causality:} While EBMs provide interpretable relationships between quality metrics and performance, they do not establish causal effects. The observed associations should be viewed as evidence of conditional patterns, not deterministic rules.
\end{itemize}

\section{Conclusion and Future Work}
\label{conclusion}
This study addressed a persistent gap in Software Defect Prediction: most prior work has examined data quality issues in isolation, focusing on class imbalance or irrelevant features while overlooking how multiple issues co-occur and interact. Such reductionist designs fail to capture the multi-problem reality of SDP datasets. To fill this gap, we systematically analyzed 374 diverse datasets using five widely adopted classifiers, quantifying five key issues—class imbalance, class overlap, irrelevant features, attribute noise, and outliers—through Explainable Boosting Machines and stratified interaction analysis.

Our findings advance the empirical basis of defect prediction in several ways. First, we show that co-occurrence is nearly universal: even the least frequent issue (attribute noise) appeared alongside others in over 93\% of datasets. Second, we identify consistent tipping points—overlap around 0.20, imbalance around 0.65–0.70, and irrelevance around 0.94—beyond which most classifiers begin to degrade. Third, we demonstrate that data quality issues interact in non-additive and sometimes counterintuitive ways: for instance, outliers, typically assumed harmful, improved performance in contexts such as high-overlap SVM or low-irrelevance DT/RF. Collectively, these results reveal a performance–robustness trade-off, showing that no single model is universally resilient.

The implications are threefold. For practitioners, the findings imply that SDP should be treated as a data-aware task: model choice and preprocessing must be conditioned on dataset-specific quality profiles, not global benchmarks. For researchers, the results imply that reporting dataset quality, modeling interactions, and including robustness metrics are essential to ensure valid comparison and theoretical progress. For the broader data-centric ML community, the findings reinforce that robustness depends less on a single “best” model and more on aligning model choice with dataset-specific quality profiles, underscoring the importance of data profiling as a first-class component of ML pipelines.

Future work can extend this agenda in several directions. First, replication with tuned and cost-sensitive configurations is needed to distinguish algorithmic limitations from artefacts of defaults. Second, expansion to emerging learners (e.g., deep neural networks, graph-based predictors) and to alternative performance metrics (MCC, F-measure, Youden's J) would broaden external validity. Finally, the development of lightweight profilers and decision aids that operationalize our thresholds offers a direct path to tool support for practitioners. 

In sum, this study advances defect prediction research by reframing data quality as a multi-dimensional, interacting phenomenon rather than a series of isolated challenges. By exposing thresholds, conditional effects, and trade-offs across models, it contributes both conceptual clarity and practical guidance, paving the way for a more data-aware and robust era of software defect prediction and, more broadly, for data-centric machine learning.


%

\section*{Declarations}

\subsection*{Funding}
Not applicable.  

\subsection*{Ethical Approval}
This study did not involve human participants or animals; ethical approval was therefore not required.  

\subsection*{Informed Consent}
Not applicable.  

\subsection*{Author Contributions}
Emmanuel Charleson Dapaah designed the study, performed the data collection and analysis, and prepared the initial draft of the manuscript.  
Jens Grabowski supervised the research and contributed to reviewing and editing the manuscript.  
Both authors read and approved the final version of the manuscript.  

\subsection*{Data Availability Statement}
The datasets and scripts generated and analyzed during this study are available in the replication package at: \\ 
\url{https://github.com/ecdapaah-dev/When-Data-Issues-Collide.git}  

\subsection*{Conflict of Interest}
The authors declare that they have no conflict of interest.  

\subsection*{Clinical Trial Number}
Not applicable.

\bibliographystyle{spbasic}      

\bibliography{mybibfile}

\end{document}